\magnification 1200
\baselineskip 15 pt

\def \v {\varphi}
\def \z {\zeta}
\def \b {\beta}
\def \P {\Phi}
\def \l {\lambda}
\def \la {\lambda}
\def \a {\alpha}
\def \ul {\underline \lambda}
\def \g {\gamma}

\def \sect#1{\bigskip  \noindent{\bf #1} \medskip }
\def \subsect#1{\bigskip \noindent{\it #1} \medskip}
\def \th#1#2{\medskip \noindent {\bf Theorem #1.}   \it #2 \rm \medskip}
\def \prop#1#2{\medskip \noindent {\bf Proposition #1.}   \it #2 \rm \medskip}
\def \cor#1#2{\medskip \noindent {\bf Corollary #1.}   \it #2 \rm \medskip}
\def \pf {\noindent  {\it Proof}.\quad }
\def \lem#1#2{\medskip \noindent {\bf Lemma #1.}   \it #2 \rm \medskip}

\def\sqr#1#2{{\vcenter{\vbox{\hrule height.#2pt\hbox{\vrule width.#2pt height#1pt \kern#1pt\vrule width.#2pt}\hrule height.#2pt}}}}

\def \square{\hfill\mathchoice\sqr56\sqr56\sqr{4.1}5\sqr{3.5}5}

\centerline{\bf Financial Valuation of Mortality Risk via the Instantaneous Sharpe Ratio:} \medskip
\centerline{\bf Applications to Pricing Pure Endowments} \bigskip

\centerline {Version: 4 November 2005}
\bigskip

\noindent Moshe A. Milevsky \hfill \break
\indent Schulich School of Business \hfill \break
\indent York University \hfill \break
\indent Toronto, Ontario, M3J 1P3 \hfill \break
\indent milevsky@yorku.ca

\medskip

\noindent S. David Promislow \hfill \break
\indent Department of Mathematics and Statistics \hfill \break
\indent York University \hfill \break
\indent Toronto, Ontario, M3J 1P3 \hfill \break
\indent promis@yorku.ca

\medskip

\noindent Virginia R. Young \hfill \break
\indent Department of Mathematics \hfill \break
\indent University of Michigan \hfill \break
\indent Ann Arbor, Michigan, 48109 \hfill \break
\indent vryoung@umich.edu

\bigskip \bigskip

\noindent{\bf Abstract:}  We develop a theory for pricing non-diversifiable mortality risk in an incomplete market. We do this by assuming that the company issuing a mortality-contingent claim requires compensation for this risk in the form of a pre-specified instantaneous Sharpe ratio.  We prove that our ensuing valuation formula satisfies a number of desirable properties.  For example, we show that it is subadditive in the number of contracts sold.  A key result is that if the hazard rate is stochastic, then the risk-adjusted survival probability is greater than the physical survival probability, even as the number of contracts approaches infinity.

\bigskip


\medskip


\medskip


\vfill \eject

\sect{1. Introduction}

Insurance markets are incomplete for several reasons, and in this paper we focus on the incompleteness arising from two sources:  (1) The timing of insurance payments is generally determined by a jump process -- such as, when individuals die or when cars crash, and (2) insurers cannot buy and sell insurance contracts frictionlessly, if at all.  Actuaries traditionally assume that one can eliminate the uncertainty associated with the jump process by selling a large number of insurance contracts.  By invoking the law of large numbers, actuaries thereby replace the random occurrence of insurance payments with a deterministic schedule.

This assumption is the reason that Brennan and Schwartz (1976), or more recently Boyle and Hardy (2003), were able to extend the Black-Scholes formula to price a derivative instrument with a random maturity date.  They assumed that the issuer sold a sufficiently large number of policies so that the issuer only had to contend with a deterministic schedule of maturing derivatives.  This simplifying technique permeates most of the literature that combines finance and insurance.  While each individual's payment occurs at a random time, one assumes that a large portfolio becomes deterministic and, thus, safe from a mortality point of view.

Using the language of modern portfolio theory, under this assumption, the idiosyncractic risk -- that is, the standard deviation per contract -- goes to zero if the insurer sells enough contracts, and so the risk is diversifiable.  However, because the insurer can only sell a finite number of insurance policies, it is impossible for the insurer to eliminate the risk that the experience will differ from what is expected.  The risk associated with selling a finite number of insurance contracts is what we call the {\it finite portfolio risk}.

In addition to the finite portfolio risk, there is a risk arising from the fact that the parameters underlying the claim occurrence process are random themselves.  For example, for an insurance payment that is contingent on the survival or death of an individual, one models the claim occurrence process as a Poisson process with intensity $\la$, also called the hazard rate.  When the individual dies, then the process jumps and is ``killed'' at that time.  Now, if $\la$ is a deterministic function of time, then the risk associated with selling a finite number of insurance contracts, each of whose claim occurrence is dictated by independent and identically distributed killed Poisson processes, is the finite portfolio risk.  If $\la$ is stochastic, then there is an additional risk that we call the {\it stochastic mortality risk}, a special case of stochastic parameter risk.  Even as the insurer sells an arbitrarily large number of contracts, the systematic stochastic mortality risk remains because of the second source of incompleteness, namely, the inability to buy and sell insurance contracts frictionlessly.

One could also define similar risks associated with the size of insurance payments.  However, for concreteness in this paper, we focus on the rate of claim occurrence and leave considerations concerning the size of the payments to future research.  In particular, we determine how to value the risks associated with a specific contract called a {\it pure endowment}.  A pure endowment pays \$1 to an individual at a given time $T$ if the individual survives to that time.  Pure endowments are important in that they are the building blocks of life annuities.  They are also analogous to defaultable zero-coupon bonds because of the uncertainty of the payment at time $T$.

We emphasize that stochastic parameter risk is different from model specification risk.  For stochastic parameter risk, we assume that we have specified the model precisely, including the stochastic process that the model parameter follows.  However, if one cannot buy and sell insurance contracts frictionlessly, one cannot eliminate the systematic risk inherent in this stochastic process.  On the other hand, if an insurer were to issue a pure endowment contract to infinitely many individuals, each of whose independent time-of-death follows the same stochastic hazard rate, then the insurer could completely hedge the stochastic mortality risk by selling term life insurance to infinitely many individuals, each of whose independent mortality follows the same stochastic hazard rate as the group buying the pure endowment.  This remark is the large-scale version of the statement that if one could sell both a pure endowment and a term life insurance contract to the same person (with the same maturity date and same mortality-contingent payment), then there would be no risk.  In this case, the price for the combined contract would be the price for a default-free zero-coupon bond.

In this paper, we consider the case in which the issuer of a pure endowment {\it cannot} sell term life insurance to hedge the stochastic mortality risk.  (In future work, we allow the issuer to hedge the risk partially by selling life insurance to an individual whose stochastic mortality is correlated with that of the purchaser of the pure endowment.)  We argue that survival probabilities are uncertain and that this uncertainty is correlated across individuals in a population -- mostly due to medical breakthroughs or environmental factors that affect the entire population.  For example, if there is a positive probability that medical science will find a cure for cancer during the next thirty years, this will influence aggregate mortality patterns.  The uncertainty regarding the evolution of survival probabilities induces a mortality dependence that cannot be diversified by selling more contracts.  This risk induces a mortality risk premium that should be priced by the market and whose magnitude depends on a representative investor's risk aversion or demanded compensation for risk.  

We are not the first to recognize that mortality rates themselves should be viewed as stochastic.  Biffis (2005), Schrager (2005), Dahl (2004), as well as Milevsky and Promislow (2001) and Soininen (1995), used and calibrated diffusion processes to model the force of mortality.  Also, demographers and actuaries, such as Lee and Carter (1992), Olivieri (2001), and DiLorenzo and Sibillo (2003), developed methods for projecting mortality.  While some practitioners, such as Smith, Moran, and Walczak (2003), and academics, such as Cox and Lin (2004) and Cairns, Blake, and Dowd (2004), implicitly recognize that mortality risk is being priced by the market, they did not derive the actual value of this risk from first principles.  In related work, Blanchett-Scalliet, El Karoui, and Martellini (2005) value assets that mature at a random time by using the principle of no arbitrage; the resulting pricing rule is, therefore, linear.  However, for insurance markets, one cannot assert that no arbitrage holds, so we take a different approach to valuing insurance contracts. 

We value a pure endowment by assuming that the insurance company issuing the contract is compensated for risk via the so-called {\it instantaneous Sharpe ratio} of a suitably-defined portfolio.  Specifically, we assume that the insurance company picks a target ratio of expected excess return to standard deviation, denoted by $\a$, and then determines a price for a pure endowment that yields this pre-determined $\a$ for the corresponding portfolio.  Our results do not depend on using a particular diffusion for the hazard rate.  In future research, we plan to calibrate this model to a generalized mean-reverting process for mortality rates, similar to the work by Norberg (2004) in the context of interest rates.


Our methodology recovers a number of results that one expects within the context of insurance, but we also obtain new insights into the breakdown of traditional insurance pricing.  For example, we prove that if the hazard rate is deterministic, then as the number of contracts approaches infinity, the price of a pure endowment collapses to the discounted expected payment using the physical probability measure to value the mortality risk, regardless of the target value of the Sharpe ratio. In other words, if the stochastic mortality risk is not present, then the price for a large portfolio reflects this and reduces to the ``usual'' expected value pricing rule in the limit.

A key result of the paper is that if the hazard rate is stochastic, then the financial value of the pure endowment is greater than the above-mentioned discounted expected payment, even as the number of contracts approaches infinity.  Furthermore, our valuation operator is subadditive and satisfies a number of other appealing properties.  Finally, our methodology allows us to decompose the value of any portfolio of pure endowment policies into a systematic component (due to uncertain aggregate mortality) and a non-systematic component (due to insuring a finite number of policies); see equation (4.55).

The remainder of this paper is organized as follows. In Section 2, we present our financial market, describe how to use the instantaneous Sharpe ratio to price the pure endowment, and derive the resulting partial differential equation that the price solves.  In Section 3, we discuss qualitative properties of the risk-adjusted price from Section 2 and show that it shares many properties with the (static) standard deviation premium principle (Gerber, 1979).  In Section 4, we study properties of the price for $n$ conditionally independent and identically distributed pure endowment risks.  In particular, in Theorem 4.11, we show that the price is subadditive with respect to $n$, and in Theorem 4.13, we show that the risk charge per person decreases as $n$ increases.  We also prove that if the hazard rate is deterministic, then the risk charge per person goes to zero as $n$ goes to infinity (Theorem 4.20 and Corollary 4.21).  Moreover, we prove that if the hazard rate is stochastic, then the risk charge person is positive as $n$ goes to infinity, which reflects the fact that the mortality risk is not diversifiable (Theorem 4.20 and Corollary 4.22).  Section 5 concludes the paper.

\sect{2. Instantaneous Sharpe Ratio}

In this section, we describe a pure endowment contract and present the financial market in which the issuer of this contract invests.  We obtain the hedging strategy for the issuer of the pure endowment.  We describe how to use the instantaneous Sharpe ratio to price the pure endowment and derive the resulting partial differential equation that the price solves.  In Sections 3 and 4, we determine properties of the price.

\subsect{2.1. Mortality Model and Financial Market}

We begin with a stochastic model for mortality.  We assume that the hazard rate $\la$ (or force of mortality) of an individual follows a diffusion process such that if the process begins at $\la_0 > \ul$ for some positive constant $\ul$, then $\la_t > \ul$ for all $t \in [0, T]$.  Thus, we require that the volatility of $\la$ goes to zero as $\la \rightarrow \ul$ from the right, and we require that the drift of $\la$ is positive for $\la$ close to $\ul$.  The reason for requiring $\la$ to have a positive lower bound $\ul$ will be apparent later.  From a modeling standpoint, $\ul$ could represent the hazard rate remaining (say, from accidents) after all biological causes of death have been removed.

Specifically, we assume that

$$d\la_t = a(\la_t, t) dt + b(t) (\la_t - \ul) dW^\la_t, \eqno(2.1)$$

\noindent in which $W^\la$ is a standard Brownian motion on a probability space $(\Omega, {\cal F}, {\bf P})$.  The volatility $b$ is either identically zero, or it is a continuous function of time $t$ bounded below by a positive constant $\kappa$ on $[0, T]$.  The drift $a$ is a H\"older continuous function of $\la$ and $t$ for which there exists $\epsilon > 0$ such that if $0 < \la - \ul < \epsilon$, then $a(\la, t) > 0$ for all $t \in [0, T]$.  After Lemma 3.3 below, we add additional requirements for $a$.  Note that if $b \equiv 0$, then $\la$ is deterministic, and in this case, we write $\l(t)$ to denote the deterministic hazard rate at time $t$.

Suppose an insurer issues a pure endowment to an individual that pays 1 at time $T$ if the individual is alive at that time.  In Section 2.2, to determine the value of the pure endowment, we will create a portfolio composed of the obligation to pay this pure endowment and of default-free zero-coupon bonds that pay 1 at time $T$ regardless of the state of the individual.  Therefore, we require a model for bond prices, and we use a model based on the short rate and the bond market's price of risk.

The dynamics of the short rate $r$, which is the rate at which the money market increases, are given by

$$dr_t = \mu(r_t, t) dt + \sigma(r_t, t) dW_t, \eqno(2.2)$$

\noindent in which $\mu$ and $\sigma \ge 0$ are deterministic functions of the short rate and time, and $W$ is a standard Brownian motion with respect to the probability space $(\Omega, {\cal F}, {\bf P})$, independent of $W^\la$.  We assume that $\mu$ and $\sigma$ are such that $r \ge 0$ almost surely.

From the principle of no-arbitrage in the bond market, there is a market price of risk $q$ for the bond that is adapted to the filtration generated by $W$; see, for example, Lamberton and Lapeyre (1996) or Bj\"ork (1998).  Moreover, the bond market's price of risk at time $t$ is a deterministic function of the short rate and of time, that is, $q_t = q(r_t, t)$.  Thus, the time $t$ price of a $T$-bond is given by

$$F(r, t; T) = {\bf E^Q} \left[ e^{-\int_t^T r_s ds} \Bigg | r_t = r \right], \eqno(2.3)$$

\noindent in which $\bf Q$ is the probability measure with Radon-Nikodym derivative with respect to $\bf P$ given by

$${ d{\bf Q} \over d{\bf P}} = e^{-\int_0^T q(r_s, s) dW_s - {1 \over 2} \int_0^T q^2(r_s, s) ds}. \eqno(2.4)$$

\noindent  It follows that $W^Q$, with $W^Q_t = W_t + \int_0^t q(r_s, s) ds$, is a standard Brownian motion with respect to $\bf Q$.

From Bj\"ork (1998), we know that the bond price $F$ solves the following partial differential equation (pde):

$$F_t + \mu^Q(r, t) F_r + {1 \over 2} \sigma^2(r, t) F_{rr} - r F = 0, \quad F(r, T; T) = 1, \eqno(2.5)$$

\noindent in which  $\mu^Q = \mu - q \sigma$.  In this paper, the horizon $T$ is fixed, so henceforth we drop $T$ from the notation of $F$.  We can use this pde to obtain the dynamics of the bond price $F(r_s, s)$, in which we think of $r_t = r$ as given and $t \le s \le T$.  Indeed,

$$\left\{ \eqalign{dF(r_s, s) &= (r_s F(r_s, s) + q(r_s, s) \sigma(r_s, s) F_r(r_s, s)) ds + \sigma(r_s, s) F_r(r_s, s) dW_s, \cr
F(r_t, t) &= F(r, t).} \right. \eqno(2.6)$$

As an aside, we could use other models commonly used in the literature and obtain the same conclusion that we reach after equation (2.17), namely, that we can factor the $T$-bond price from the ``mortality price.''  We use a model involving the short rate and the bond market's price of risk for ease of presentation.

\subsect{2.2. Pricing via the Instantaneous Sharpe Ratio}

The insurer faces the unhedgeable risk that the individual's living or dying will be different from expected; therefore, the insurer demands a return greater than the sum of the return $r$ on the money market and the ``return'' $\la$ from the mortality component.  One measure of the risk that the insurer takes is the standard deviation of the change in the portfolio. A natural tie between the excess return and the standard deviation is the ratio of the former to the latter, the so-called {\it instantaneous Sharpe ratio}.  In what follows, we find the hedging strategy to minimize the local variance of the change in the portfolio, then we set the price of the pure endowment so that the resulting instantaneous Sharpe ratio equals a given constant.  We could set the instantaneous Sharpe ratio equal to a function of $\la$ and $t$, but we choose a constant for simplicity.

The market for insurance is incomplete; therefore, there is no unique pricing mechanism.  To value contracts in this market, one must assume something about how risk is ``priced.''  For example, one could use the principle of equivalent utility (see Zariphopoulou (2001) for a review) or the Esscher transform (Gerber and Shiu, 1994) to price the risk.  We employ the instantaneous Sharpe ratio because of its analogy with the bond market's price of risk and because of the desirable properties of the resulting price.  Because of these properties, we anticipate that our pricing methodology will prove useful in pricing risks in other incomplete markets. 

Denote the value (price) of the pure endowment by $P = P(r, \la, t)$, in which we explicitly recognize that the price of the pure endowment will depend on the short rate $r$ and the hazard rate $\la$ at time $t$.  Suppose the insurer creates a portfolio $\Pi$ with value $\Pi_t$ at time $t$.  The portfolio contains the obligation to pay the pure endowment at time $T$ if the individual is alive at that time, namely $-P$.  Additionally, the insurer holds $\pi_t$ $T$-bonds.  Thus, $\Pi_t = -P(r_t, \la_t, t) + \pi_t F(r_t, t)$.

By It\^o's Lemma (Protter, 1995), the value of the portfolio at time $t + h$ with $h > 0$, namely $\Pi_{t + h}$, equals

$$\eqalign{\Pi_{t+h} &= \Pi_t - \int_t^{t+h} {\cal D}^\mu P(r_s, \la_s, s) ds + \int_t^{t+h} \sigma(r_s, s) (\pi_s F_r(r_s, s) - P_r(r_s, \la_s, s)) \, dW_s \cr
& \quad - \int_t^{t+h} b(s) (\la_s - \ul) P_\la(r_s, \la_s, s) \, dW^\la_s + \int_t^{t+h} P(r_s, \la_s, s) (dN_s - \la_s \, ds)  \cr
& \quad + \int_t^{t+h}  \pi_s (r_s F(r_s, s) + q(r_s, s) \sigma(r_s, s) F_r(r_s, s)) \, ds,} \eqno(2.7)$$

\noindent in which ${\cal D}^m$, with $m = m(r, t)$ a deterministic function of the short rate and time, is an operator defined on the set of appropriately differentiable functions on ${\bf R}^+ \times (\ul, \infty) \times [0, T]$ by

$${\cal D}^m v = v_t + m v_r + {1 \over 2} \sigma^2 v_{rr} + a v_\la + {1 \over 2} b^2 (\la - \ul)^2 v_{\la \la} - \la v.  \eqno(2.8)$$

\noindent Also, in (2.7), $N$ denotes a Poisson process with stochastic parameter $\la$.  Thus, $\Pi$ jumps in value by $P$ when an individual dies because the insurer is no longer responsible for paying 1 at time $T$.

In this single-life case, the process $\Pi$ is ``killed'' when the individual dies.  If we were to consider the price $P^{(n)}$ for $n$ conditionally independent and identically distributed lives (conditionally independent given the hazard rate), then $N$ would be a Poisson process with stochastic parameter $n \la$ such that $\Pi$ jumps by $P^{(n)} - P^{(n - 1)}$ when an individual dies.  We consider $P^{(n)}$ later and continue with the single-life case now.

We next calculate the expectation and variance of $\Pi_{t+h}$ conditional on the information available at time $t$, namely ${\cal F}_t$.  First,

$$\eqalign{{\bf E}(\Pi_{t+h} | {\cal F}_t) &= \Pi - {\bf E}^{r, \la, t} \int_t^{t+h} {\cal D}^\mu P(r_s, \la_s, s) ds \cr
& \quad + {\bf E}^{r, \la, t} \int_t^{t+h} \pi_s (r_s F(r_s, s) + q(r_s, s) \sigma(r_s, s) F_r(r_s, s)) \, ds.}  \eqno(2.9)$$

\noindent Here, $\Pi_t = \Pi$ is known at time $t$, and ${\bf E}^{r, \la, t}$ denotes the conditional expectation given $r_t = r$ and $\la_t = \la$.  Define the stochastic process $Y_h$ for $h \ge 0$ by

$$Y_h = \Pi - \int_t^{t+h} {\cal D}^\mu P(r_s, \la_s, s) ds + \int_t^{t+h} \pi_s (r_s F(r_s, s) + q(r_s, s) \sigma(r_s, s) F_r(r_s, s)) ds.  \eqno(2.10)$$

\noindent Thus, ${\bf E}(\Pi_{t+h} | {\cal F}_t) = {\bf E}^{r, \la, t} Y_h$, and from (2.7), we have

$$\eqalign{\Pi_{t + h} &= Y_h + \int_t^{t+h} \sigma(r_s, s) (\pi_s F_r(r_s, s) - P_r(r_s, \la_s, s)) \, dW_s \cr
& \quad - \int_t^{t+h} b(s) (\la_s - \ul) P_\la(r_s, \la_s, s) \, dW^\la_s + \int_t^{t+h} P(r_s, \la_s, s) (dN_s - \la_s \, ds).} \eqno(2.11)$$

\noindent It follows that

$$\eqalign{& {\bf Var}(\Pi_{t+h} | {\cal F}_t) = {\bf E}((\Pi_{t+h} - {\bf E}Y_h)^2 | {\cal F}_t) \cr
& \quad = {\bf E}^{r, \la, t} (Y_h - {\bf E}Y_h)^2 + {\bf E}^{r, \la, t} \int_t^{t+h} \sigma^2(r_s, s) (\pi_s F_r(r_s, s) - P_r(r_s, \la_s, s))^2 ds \cr
& \qquad + {\bf E}^{r, \la, t}  \int_t^{t+h} b^2(s) (\la_s - \ul)^2 P_\la^2(r_s, \la_s, s) ds + {\bf E}^{r, \la, t}  \int_t^{t+h} \la_s P^2(r_s, \la_s, s) ds.} \eqno(2.12)$$

We choose $\pi_t$ in order to minimize the local variance $\lim_{h \rightarrow 0} {1 \over h} {\bf Var}(\Pi_{t+h} | {\cal F}_t)$, a dynamic measure of risk of the portfolio; therefore, $\pi_t = P_r(r_t, \la_t, t)/F_r(r_t, t)$.  Under this assignment, the drift and local variance become, respectively,

$$\lim_{h \rightarrow 0} {1 \over h} ({\bf E} (\Pi_{t+h} | {\cal F}_t) - \Pi) = - {\cal D}^{\mu^Q} P(r, \la, t) + r P_r(r, \la, t) {F(r, t) \over F_r(r, t)},  \eqno(2.13)$$

\noindent and

$$\lim_{h \rightarrow 0} {1 \over h} {\bf Var}(\Pi_{t+h} | {\cal F}_t) = b^2(t) (\la - \ul)^2 P_\la^2(r, \la, t) + \la P^2(r, \la, t). \eqno(2.14)$$

Now, we come to pricing via the instantaneous Sharpe ratio.  Because the minimum local variance  in (2.14) is positive, the insurer is unable to completely hedge the risk of the pure endowment contract.  Therefore, the price should reimburse the insurer for its risk, say, by a constant multiple $\a$ of the local standard deviation of the portfolio.  It is this $\a$ that is the instantaneous Sharpe ratio.

From (2.14), we learn that the local standard deviation of the portfolio equals

$$\lim_{h \rightarrow 0} \sqrt{{1 \over h}  {\bf Var}(\Pi_{t+h} | {\cal F}_t)} = \sqrt{b^2(t)(\la - \ul)^2 P^2_\la(r, \la, t) + \la P^2(r, \la, t)}.  \eqno(2.15)$$

\noindent To determine the value (price) $P$, we set the drift of the portfolio equal to the short rate times the portfolio {\it plus} $\a$ times the local standard deviation.  Thus, from (2.13) and (2.15), we have that $P$ solves the equation

$$- {\cal D}^{\mu^Q} P + r P_r {F \over F_r} = r \Pi + \a \sqrt{b^2(t)(\la - \ul)^2 P_\la^2 + \la P^2}, \eqno(2.16)$$

\noindent for some $0 \le \a \le \sqrt{\ul}$.  Recall that $\Pi = -P + \pi F = -P + P_r F/F_r$.  It follows that $P = P(r, \la, t)$ solves the non-linear pde given by

$$\left\{ \eqalign{&P_t + \mu^Q P_r + {1 \over 2} \sigma^2 P_{rr} + a P_\la + {1 \over 2} b^2 (\la - \ul)^2 P_{\la \la} - (r + \la)P \cr
& \quad = - \a \sqrt{b^2 (\la - \ul)^2 P_\la^2 + \la P^2} \cr
& P(r, \la, T) = 1.} \right. \eqno(2.17)$$

If we had been able to choose the investment strategy $\pi$ so that the local standard deviation in (2.15) were identically zero (that is, if the risk were hedgeable), then the right-hand side of the pde in (2.17) would be zero, and we would have a linear differential equation of the Black-Scholes type.  One can think of the right-hand side as adding a margin to the return of the portfolio because the pure endowment risk is not completely hedgeable due to the mortality risk.  In addition to the unhedgeable mortality risk, the insurer also faces the somewhat diversifiable finite portfolio risk, that is, the risk that even if the hazard rate is deterministic, the actual number who survive until time $T$ is different from expected.  This risk is clearly present when selling a pure endowment to a single individual, but it is also present to some extent in any portfolio of finite size.  In Section 4.4, we decompose the risk loading in the price due to the finite portfolio risk and due to the stochastic mortality risk. 

If there were no risk loading, that is, if $\a = 0$, then the price is such that the expected return on the price is $r + \la$.  The rate $r$ arises from the riskless money market, and $\la$ arises from the expected release of reserves as individuals die.  If $\a > 0$, then the expected return on the price is greater than $r + \la$.  Therefore, $\a$, the Sharpe ratio, measures the degree to which the insurer's total expected return is in excess of $r + \la$, as a proportion of the standard deviation of the return.

Before moving on to the following sections where we study properties of the solution of (2.17), we show that we can simplify $P$ greatly.  Indeed, $P(r, \la, t) = F(r, t) \v(\la, t)$, in which $F$ is the price of the $T$-bond and solves (2.5), and $\v$ solves the non-linear pde

$$\left\{ \eqalign{& \v_t + a \v_\la +  {1 \over 2} b^2 (\la - \ul)^2 \v_{\la \la} - \la \v = - \a \sqrt{b^2 (\la - \ul)^2 \v^2_\la + \la \v^2}, \cr
& \v(\la, T) = 1.} \right. \eqno(2.18)$$

\noindent The existence of a solution to (2.18) follows from standard techniques; see, for example, Walter (1970, Chapter IV, Section 36).  A comparison result (see Section 3 of this paper) demonstrates that the solution is unique.

The factorization $P = F \v$ is reminiscent of the standard actuarial method of pricing pure endowments in that $\v$ represents the probability of paying the mortality-contingent claim, that is, the probability that the individual survives.  We will see in Section 3.1 that we can interpret $\v$ as a risk-adjusted survival probability.  As mentioned at the end of Section 2.1, this factorization arises under other commonly used models for bond prices, as long as the risk driving the bond price ($W$ in our case) is independent of the risk driving the stochastic hazard rate, $W^\la$. 

Consider the special case for which $b \equiv 0$, that is, $\la$ is deterministic.  Suppose $\l(t)$ is the solution of $d \la = a(\la, s) ds$ with initial value $\la_0 = \la$; then, (2.18) becomes the linear ordinary differential equation

$$\v'(t) - (\l(t) - \a \sqrt{\l(t)}) \v(t) = 0, \qquad \v(T) = 1, \eqno(2.19)$$

\noindent whose solution is

$$\v(t) = e^{ - \int_t^T (\l(s) - \a \sqrt{\l(s)} \, ) ds} \, , \eqno(2.20)$$

\noindent a type of probability of survival because we can think of $\l(t) - \a \sqrt{\l(t)} > 0$ as a modified hazard rate.  Indeed, note that if $\a = 0$, then (2.20) is the physical probability that a person alive at time $t$ survives to time $T$, and as $\a$ increases, $\v$ increases.  Therefore, we interpret (2.20) as a risk-adjusted probability of survival, in which $\a$ controls the degree to which we adjust (that is, increase) the physical probability of survival.  When $b \not\equiv 0$, a similar phenomenon occurs (see Theorem 3.8 below), and for this reason, we refer to the solution $P$ of (2.17) as the risk-adjusted price for the pure endowment.

Recall that we assume that $0 \le \a \le \sqrt{\ul}$.  For the case of deterministic hazard, this implies that $\l(t) - \a \sqrt{\l(t)} > 0$ for all $t \in [0, T]$, so that $0 \le \v(t) \le 1$ in (2.20).  For stochastic hazard, we show in the next section that $0 \le \v(\la, t) \le 1$ for all $(\la, t) \in (\ul, \infty) \times [0, T]$.  It follows that, in general, $F$ is an upper bound for the price.  Observe that $F$ is a natural upper bound for the price because it is the price we would charge if we knew the person could not die before $T$.

\sect{3. Qualitative Properties of the Risk-Adjusted Price}

In this section, we discuss qualitative properties of the risk-adjusted price $P$ in (2.17) and show that it shares many properties with the (static) standard deviation premium principle (Gerber, 1979).  To begin, we have the following proposition.

\prop{3.1}{Suppose $P^c$ is the price, as determined by the method in Section 2, for a pure endowment with payment $c \ge 0$ at time $T$ if the individual is alive.  Then, $P^c = c P$, in which $P$ is the risk-adjusted price for a payment of 1 at time $T$ if the individual is alive.}

\pf  In the derivation of $P$ for (2.17), it is clear that if we derive the price $P^c$, then (2.17) still applies for determining $P^c$ with the terminal condition $P^c(r, \la, T) = c$.  We can, then, write $P^c = c F \v = cP$.  $\square$

\medskip

Proposition 3.1 parallels the following well known fact concerning the (static) standard deviation premium principle $H$:  If we define the standard deviation premium principle, as applied to a random variable $X$, by

$$H(X) = {\bf E}X + \alpha \sqrt{{\bf Var}X}, \eqno(3.1)$$

\noindent then $H(c X) = c H(X)$ for $c \ge 0$.

In what follows, we show that $0 \le P \le F$ and $P_\la \le 0$, and we examine how the price $P$ responds to changes in the model parameters.  To this end, we need a comparison principle (Walter, 1970, Section 28).  We begin by stating a relevant one-sided Lipschitz condition along with growth conditions.  We require that the function $g = g(\la, t, v, p)$ satisfies the following one-sided Lipschitz condition:  For $v > w$,

$$g(\la, t, v, p) - g(\la, t, w, q) \le c(\la, t) (v - w) + d(\la, t) |p - q|, \eqno(3.2)$$

\noindent with growth conditions on $c$ and $d$ given by

$$0 \le c(\la, t) \le K(1 + (\ln(\la - \ul))^2), \hbox{ and } 0 \le d(\la, t) \le K (\la - \ul) (1 + |\ln(\la - \ul)|), \eqno(3.3)$$

\noindent for some constant $K \ge 0$, and for all $(\la, t) \in (\ul, \infty) \times [0, T]$.  Throughout this paper, we rely on the following useful comparison principle, which we obtain from Walter (1970, Section 28).

\th{3.2} {Let $G = (\ul, \infty) \times [0, T],$ and denote by $\cal G$ the collection of functions on $G$ that are twice-differentiable in their first variable and once-differentiable in their second. Define a differential operator $\cal L$ on $\cal G$ by
$${\cal L} v = v_t + {1 \over 2} b^2(t) (\la - \ul)^2 v_{\la \la} + g(\la, t, v, v_\la),  \eqno(3.4)$$
\noindent in which $g$ satisfies $(3.2)$ and $(3.3)$.  Suppose $v, w \in \cal G$ are such that there exists a constant $K \ge 0$ with $v \le e^{K (\ln(\la - \ul))^2}$ and $w \ge - e^{K (\ln(\la - \ul))^2}$ for large $\la$ and for $\la$ close to $\ul$. Then, if $($a$)$ ${\cal L} v \ge {\cal L} w$ on $G,$ and if $($b$)$ $v(\la, T) \le w(\la, T)$ for all $\la > \ul$, then $v \le w$ on $G$.}

\medskip

\pf Transform the variables $\la$ and $t$ in (3.4) to $y = \ln(\la - \ul)$ and $\tau = T - t$, and write $\tilde v(y, \tau) = v(\la, t)$, etc.  Under this transformation, (3.4) becomes

$${\cal L} \tilde v = - \tilde v_\tau + {1 \over 2} \tilde b^2(\tau) \tilde v_{yy} + \tilde h(y, \tau, \tilde v, \tilde v_y),  \eqno(3.5)$$

\noindent in which $\tilde h(y, \tau, \tilde v, \tilde p) = - {1 \over 2} \tilde b^2(\tau) \tilde p + \tilde g(y, \tau, \tilde v, \tilde p)$, and $\tilde v$ is a differential function on ${\bf R} \times [0, T]$.  Note that $\v_\la = e^{-y} \tilde \v_y$, so $p = e^{-y} \tilde p$ in going from $g$ to $\tilde g$.  The differential operator in (3.5) is of the form considered by Walter (1970, Section 28, pages 213-215); see that reference for the proof of our assertion.

The remaining item to consider is the form of the growth conditions in the original variables $\la$ and $t$.  From Walter (1970), we know that analog of (3.2) and (3.3) for $\tilde h$ are

$$\tilde h(y, \tau, \tilde v, \tilde p) - \tilde h(y, \tau, \tilde w, \tilde q) \le \tilde c(y, \tau) (\tilde v - \tilde w) + \tilde d(y, \tau) |\tilde p -  \tilde q|, \eqno(3.6)$$

\noindent with

$$0 \le \tilde c(y, \tau) \le K(1 + y^2), \hbox{ and } 0 \le \tilde d(y, \tau) \le K (1 + |y|). \eqno(3.7)$$

\noindent Under the original variables, the right-hand side of (3.6) becomes $c(\la, t) (v - w) + d(\la, t) |p - q|$, in which $c(\la, t) = \tilde c(y, \tau)$ and $d(\la, t) = \tilde d(y, \tau) e^y$ because $\tilde p = e^y p$.  Therefore, $\tilde d(y, \tau) \le K (1 + |y|)$ becomes $d(\la, t) \le K e^y (1 + |y|) = K (\la - \ul) (1 + | \ln(\la - \ul) |)$.  $\square$

\medskip

As a lemma for results to follow, we show that the differential operator associated with our problem satisfies the hypotheses of Theorem 3.2.

\lem{3.3} {If we define $g$ by
$$g(\la, t, v, p) = a(\la, t) p - \la v + \a \sqrt{b^2(t) (\la - \ul)^2 p^2 + \la v^2}, \eqno(3.8)$$
\noindent then $g$ satisfies the one-sided Lipschitz condition $(3.2)$ on $G$.  Furthermore, if $|a(\la, t) | \le K (\la - \ul) (1 + |\ln(\la - \ul)|),$ then $(3.3)$ holds.}

\pf Suppose $v > w$,  then

$$\eqalign{&g(\la, t, v, p) - g(\la, t, w, q) = a(\la, t) (p - q) - \la(v - w) \cr
& \qquad \qquad + \a \left\{ \sqrt{b^2(t) (\la - \ul)^2 p^2 + \la v^2} - \sqrt{b^2(t) (\la - \ul)^2 q^2 + \la w^2} \right\} \cr
& \quad \le ( | a(\la, t) | + \a b(t) (\la - \ul) ) |p-q| - \left(\la - \a \sqrt{\la} \right) (v - w) \cr
& \quad \le ( | a(\la, t) | + \a b(t) (\la - \ul) ) |p-q|.} \eqno(3.9)$$

\noindent Recall that $\a \le \sqrt{\ul}$.   Also, we use the fact that if $A \ge B$, then $\sqrt{C^2 + A^2} - \sqrt{C^2 + B^2} \le A - B$, as we demonstrate below in Lemma 4.5.  Thus, (3.2) holds with $c(\la, t) = 0$ and $d(\la, t) =  | a(\la, t) | + \a b(t) (\la - \ul)$.  Note that $d$ satisfies (3.3) if $|a(\la, t) | \le K (\la - \ul) (1 + |\ln(\la - \ul)|)$.  $\square$

\medskip

\noindent{\bf Assumption 3.4.} Henceforth, we assume that the drift $a$ satisfies the growth condition in the hypothesis of Lemma 3.3.  For later purposes (for example, see Theorem 3.7), we also assume that $a_\la$ is H\"older continuous and satisfies the growth condition $|a_\la| \le K(1 + (\ln(\la - \ul))^2)$.

\medskip

In the next two subsections, we apply Theorem 3.2 and Lemma 3.3 repeatedly to determine qualitative properties of the risk-adjusted premium $P$.

\subsect{3.1. Interpreting $\v$ as a Survival Probability}

In our first applications of Theorem 3.2, we show that $\v$ shares two important properties with physical survival   probabilities, namely that $0 \le \v \le 1$ and $\v_\la \le 0$.  In fact, we have an even tighter upper bound on $\v$ by observing that $\ul$ is a lower bound for $\la$.

\th{3.5} {$0 \le \v(\la, t) \le e^{- \left(\ul - \a \sqrt{\ul} \right)(T - t)}$ for $(\la, t) \in G$.}

\pf  Define the differential operator $\cal L$ on $\cal G$ by (3.4) with $g$ given in (3.8).  Because $\v$ solves (2.18), we have ${\cal L} \v = 0$.  Also,

$$\eqalign{{\cal L} e^{- \left(\ul - \a \sqrt{\ul} \right)(T - t)} &= \left(\ul - \a \sqrt{\ul} \right) e^{- \left(\ul - \a \sqrt{\ul} \right)(T - t)} - \left(\la - \a \sqrt{\la} \right) e^{- \left(\ul - \a \sqrt{\ul} \right)(T - t)} \cr
& \propto \left(\ul - \a \sqrt{\ul} \right) - \left(\la - \a \sqrt{\la} \right) \le 0.}  \eqno(3.10)$$

\noindent Because ${\cal L} e^{- \left(\ul - \a \sqrt{\ul} \right)(T - t)} \le {\cal L} \v$ and $\v(\la, T) = 1 = e^{-\left(\ul - \a \sqrt{\ul} \right)(T - T)}$, Theorem 3.2 and Lemma 3.3 imply that $\v \le e^{-\left(\ul - \a \sqrt{\ul} \right)(T - t)}$ on $G$.

Similarly, denote by {\bf 0} the function that is identically 0 on $G$; then, ${\cal L} {\bf 0} = 0 = {\cal L} \v.$  Because additionally $\v(\la, T) = 1$, Theorem 3.2 and Lemma 3.3 imply that $0 \le \v$ on $G$.   $\square$

\medskip

The upper bound in Theorem 3.5 is tight.  Indeed, suppose $\la \equiv \ul + \epsilon$ for some constant $\epsilon > 0$.  Then, the solution to (2.18) is given by $e^{- \left((\ul + \epsilon) - \a \sqrt{\ul + \epsilon} \right)(T - t)}$, which can be made arbitrarily close to $e^{- \left(\ul - \a \sqrt{\ul} \right)(T - t)}$ (uniformly on $[0, T]$) by choosing $\epsilon$ small enough.

We have the following corollary of Theorem 3.5 that gives us natural bounds for the risk-adjusted price.  $F$ is a natural bound for the price because it is the price of a default-free bond, that is, a bond that pays regardless of whether the individual is alive.

\cor{3.6} {$0 \le P(r, \la, t) \le F(r, t)$ for $(r, \la, t) \in {\bf R}^+ \times G$.}

\pf Because  $P(r, \la, t) = F(r, t) \v(\la, t)$ and $e^{- \left(\ul - \a \sqrt{\ul} \right)(T - t)} \le 1$, the result is immediate from Theorem 3.5.  $\square$

\medskip

We end this subsection with a proof that $\v_\la \le 0$.  This result is intuitive for physical survival probabilities in that if the current hazard rate $\la$ increases, then the probability of surviving until time $T$ decreases.

\th{3.7} {$\v_\la(\la, t) \le 0$ for $(\la, t) \in G$.}

\pf  To prove this assertion, we apply a modified version of Theorem 3.2 to the special case of comparing $\v_\la$ with the zero function {\bf 0}.  From Walter (1970, Section 28, pages 213-215), we see that we only need to verify that (3.2) holds for $v > 0 = w = q$.  First, differentiate $\v$'s equation with respect to $\la$ to get an equation for $f = \v_\la$.

$$\left\{ \eqalign{& f_t + (a_\la - \la)f + (a + b^2(\la - \ul)) f_\la +  {1 \over 2} b^2 (\la - \ul)^2 f_{\la \la} - \v \cr
& \quad = - \a { b^2 (\la - \ul) f^2 + b^2 (\la - \ul)^2 f f_\la + {1 \over 2} \v^2 + \la \v f \over \sqrt{b^2 (\la - \ul)^2 f^2 + \la \v^2}}, \cr
& f(\la, T) = 0.} \right. \eqno(3.11)$$

Define a differential operator $\cal L$ on $\cal G$ by (3.4) with $g$ given by

$$g(\la, t, v, p) = (a_\la - \la)v + (a + b^2(\la - \ul)) p - \v + 
 \a { b^2 (\la - \ul) v^2 + b^2 (\la - \ul)^2 v p + {1 \over 2} \v^2 + \la \v v \over \sqrt{b^2 (\la - \ul)^2 v^2 + \la \v^2}}. \eqno(3.12)$$
 
\noindent To apply a modified version of Theorem 3.2, verify that (3.2) and (3.3) hold for $v > 0 = w = q$.  It is not difficult to show that in this case,

$$\eqalign{g(\la, t, v, p) - g(\la, t, 0, 0) &\le (| a_\la | + \a b - (\la - \a \sqrt{\la}) v + ( |a| + b^2(\la - \ul) + \a b(\la - \ul)) | p | \cr
&\le (| a_\la | + \a b) v + ( |a| + b^2(\la - \ul) + \a b(\la - \ul)) | p |.} \eqno(3.13)$$ 

\noindent Thus, by Assumption 3.4, $g$ satisfies (3.2) with the corresponding $c$ and $d$ satisfying the growth conditions in (3.3).

Next, note that because $f = \v_\la$ satisfies (3.11), ${\cal L} f = 0$.  Also, ${\cal L} {\bf 0} = \v ( -1 + \a/(2 \sqrt{\la}) )  \le 0$ because $\la \ge \ul \ge \a^2$.  These observations, together with $f(\la, T) = 0$, imply that $f = \v_\la \le 0$ on $G$.  $\square$

\medskip

As in the relationship between Theorem 3.5 and Corollary 3.6, we have the immediate corollary of Theorem 3.7 that $P_\la \le 0$ on $G$.

\subsect{3.2. Comparative Statics for $P$}

Our next results show that as we vary the model parameters, the price $P$ responds consistently with what we expect.

\th{3.8} {Suppose $0 \le \a_1 <  \a_2 \le \sqrt{\ul},$ and let $P^{\a_i}$ be the solution to $(2.17)$ with $\a = \a_i,$ for $i = 1, 2$. Then, $P^{\a_1}(r, \la, t) \le P^{\a_2}(r, \la, t)$ for all $(r, \la, t) \in {\bf R}^+ \times G$.}

\pf Because $F$ in (2.5) is independent of $\a$, it is enough to show that $\v^{\a_1} \le \v^{\a_2}$ on $G$, in which $\v^{\a_i}$ has the obvious meaning.  Define a differential operator $\cal L$ on $\cal G$ by (3.4) and (3.8) with $\a = \a_1$.  Because $\v^{\a_1}$ solves (2.18) with $\a = \a_1$, we have ${\cal L} \v^{\a_1} = 0$.  Also,

$$\eqalign{{\cal L} \v^{\a_2} &= \v^{\a_2}_t + a \v^{\a_2}_\la + {1 \over 2} b^2 (\la - \ul)^2 \v^{\a_2}_{\la \la} - \la \v^{\a_2} + \a_1 \sqrt{b^2 (\la - \ul)^2 (\v^{\a_2}_\la)^2 + \la (\v^{\a_2})^2} \cr
&= -(\a_2 - \a_1) \sqrt{b^2 (\la - \ul)^2 (\v^{\a_2}_\la)^2 + \la (\v^{\a_2})^2} \le 0 = {\cal L} \v^{\a_1}.} \eqno(3.14)$$

\noindent In addition, both $\v^{\a_1}$ and $\v^{\a_2}$ satisfy the terminal condition $\v^{\a_i}(\la, T) = 1$.  Thus, Theorem 3.2 and Lemma 3.3 imply that $\v^{\a_1} \le \v^{\a_2}$ on $G$.  $\square$

\medskip 

Theorem 3.8 states that as the parameter $\a$ increases, the price $P^\a$ increases; this result justifies the use of the phrase {\it risk parameter} when referring to $\a$.  It is clear that the price obtained from the standard deviation premium principle, as defined in (3.1), also increases with $\a$.  We have the following corollary to Theorem 3.8.

\cor{3.9} {Let $P^{\a0}$ be the solution to $(2.17)$ with $\a = 0;$ then, $P^{\a0} \le P^{\a}$ for all $0 \le \a \le \sqrt{\ul},$ and we can express the lower bound $P^{\a0}$ as follows:  $P^{\a0} (r, \la, t) = F(r, t) \v^{\a0} (\la, t),$ in which $\v^{\a0}$ is given by the physical probability of survival, namely}

$$\v^{\a0} (\la, t) = {\bf E} \left[  e^{ - \int_t^T \la_s ds} \Big| \la_t = \la \right],  \eqno(3.15)$$

\noindent {\it where $\la_s$ follows the process given in $(2.1)$.}

\medskip

\pf Theorem 3.8 implies that $P^{\a0} \le P^{\a}$ for all $0 \le \a \le \sqrt{\ul}$, and by substituting $\a = 0$ in (2.18), the Feynman-Kac Theorem (Karatzas and Shreve, 1991) implies the representation of $\v^{\a0}$ in (3.15).  $\square$

\medskip

Corollary 3.9 justifies the use of the phrase {\it risk-adjusted} price when referring to $P^\a$ because $P^\a \ge P^{\a0}$, and we can interpret $P^{\a0}$ as a {\it risk-neutral} price due to the fact that $P^{\a0}$ equals the product of the bond price and the probability of paying the pure endowment.  We call $P^\a - P^{\a0}$ the {\it risk charge} that compensates the insurer for both the finite portfolio risk and the stochastic mortality risk.  In Section 4.4, we decompose the risk charge into these two components after we study the price for a portfolio of $n$ pure endowment risks.

Next, we examine how the risk-adjusted price $P$ varies with the drift and volatility of the stochastic hazard rate.

\th{3.10} {Suppose $a_1(\la, t) \le a_2(\la, t)$ on $G$, and let $P^{a_i}$ denote the solution to $(2.17)$ with $a = a_i,$  for $i = 1, 2$.  Then, $P^{a_1}(r, \la, t) \ge P^{a_2}(r, \la, t)$ for all $(r, \la, t) \in {\bf R}^+ \times G$.}

\pf Because $F$ in (2.5) is independent of $a$, it is enough to show that $\v^{a_1} \ge \v^{a_2}$, in which $\v^{a_i}$ has the obvious meaning.  Define a differential operator $\cal L$ on $\cal G$ by (3.4) and (3.8) with $a = a_1$.  Because $\v^{a_1}$ solves (2.18) with $a = a_1$, we have ${\cal L} \v^{a_1} = 0$.  Also, because $\v_\la^{a_2} \le 0$, then

$$\eqalign{{\cal L} \v^{a_2} &= \v^{a_2}_t + a_1 \v^{a_2}_\la + {1 \over 2} b^2 (\la - \ul)^2 \v^{a_2}_{\la \la} - \la \v^{a_2} + \a \sqrt{b^2 (\la - \ul)^2 (\v^{a_2}_\la)^2 + \la (\v^{a_2})^2} \cr
&= -(a_2 - a_1) \v^{a_2}_\la \ge 0 = {\cal L} \v^{a_1}.} \eqno(3.16)$$

\noindent In addition, both $\v^{a_1}$ and $\v^{a_2}$ satisfy the terminal condition $\v^{a_i}(\la, T) = 1$. Theorem 3.2 and Lemma 3.3 imply that $\v^{a_1} \ge \v^{a_2}$ on $G$.  $\square$

\medskip

The result parallel to Theorem 3.10 for the standard deviation premium principle in (3.1) is that $H$ decreases as ${\bf E}X$ decreases.

\th{3.11} {Suppose $0 \le b_1(t) \le b_2(t)$ on $[0, T],$ and let $P^{b_i}$ denote the solution to $(2.17)$ with $b = b_i,$  for $i = 1, 2$.  If $P^{b_i}_{\la \la} \ge 0$ for $i = 1$ or $2,$ then $P^{b_1}(r, \la, t) \le P^{b_2}(r, \la, t)$ for all $(r, \la, t) \in {\bf R}^+ \times G$.}

\pf Because $F$ in (2.5) is independent of $b$, it is enough to show that $\v^{b_1} \le \v^{b_2}$ if $\v^{b_i}_{\la \la} \ge 0$, in which $\v^{b_i}$ has the obvious meaning.  Suppose $\v_{\la \la}^{b_2} \ge 0$, and define a differential operator $\cal L$ on $\cal G$ by (3.4) and (3.8) with $b = b_1$.  Because $\v^{b_1}$ solves (2.18) with $b = b_1$, we have ${\cal L} \v^{b_1} = 0$.  Also,

$$\eqalign{{\cal L} \v^{b_2} &= \v^{b_2}_t + a \v^{b_2}_\la + {1 \over 2} b_1^2 (\la - \ul)^2 \v^{b_2}_{\la \la} - \la \v^{b_2} + \a \sqrt{b_1^2 (\la - \ul)^2 (\v^{b_2}_\la)^2 + \la (\v^{b_2})^2} \cr
&= -{1 \over 2} \left(b_2^2 - b_1^2 \right) (\la - \ul)^2 \v^{b_2}_{\la \la} \cr
& \qquad - \a \left\{ \sqrt{b_2^2 (\la - \ul)^2 (\v^{b_2}_\la)^2 + \la (\v^{b_2})^2} - \sqrt{b_1^2 (\la - \ul)^2 (\v^{b_2}_\la)^2 + \la (\v^{b_2})^2} \right\} \cr
&\le 0 = {\cal L} \v^{b_1}.} \eqno(3.17)$$

\noindent In addition, both $\v^{b_1}$ and $\v^{b_2}$ satisfy the terminal condition $\v^{b_i}(\la, T) = 1$.  Thus, Theorem 3.2 and Lemma 3.3 imply that $\v^{b_1} \le \v^{b_2}$ on $G$.  In a similar fashion, we obtain the same result if $\v_{\la \la}^{b_1} \ge 0$.  $\square$

\medskip

From Theorem 3.11, we see that if $\v$ is convex with respect to $\la$, then the risk-adjusted price increases if the volatility on the stochastic hazard rate increases.  The parallel result for the standard deviation premium principle in (3.1) is that $H$ increases as ${\bf Var}X$ increases.

\medskip

\sect{4. The Risk-Adjusted Price for a Portfolio of Risks}

In this section, we study properties of the price $P^{(n)} = P^{(n)}(r, \la, t)$ for $n$ conditionally independent and identically distributed pure endowment risks.  First, we present the equation that $P^{(n)}$ solves.  In Sections 4.1 and 4.2, we parallel the results of Sections 3.1 and 3.2, respectively.  In Section 4.3, we show that $P^{(n)}$ is subadditive.  In Section 4.4, we show that the risk charge per person decreases as the $n$ increases.  We also show that if the hazard rate is deterministic, then the risk charge per person goes to zero as $n$ goes to infinity.  Moreover, we show that if the hazard rate is stochastic, then the risk charge person is positive as $n$ goes to infinity, which reflects the fact that the stochastic mortality risk is not diversifiable.

As discussed in the paragraph preceding equation (2.9), when an individual dies, the portfolio value $\Pi$ jumps by $P^{(n)} - P^{(n-1)}$.  By paralleling the derivation of (2.17), one can show that $P^{(n)}$ solves the non-linear pde given by

$$\left\{ \eqalign{&P^{(n)}_t + \mu^Q P^{(n)}_r + {1 \over 2} \sigma^2 P^{(n)}_{rr} + a P^{(n)}_\la + {1 \over 2} b^2 (\la - \ul)^2 P^{(n)}_{\la \la} - r P^{(n)} - n \la \left(P^{(n)} - P^{(n-1)} \right) \cr
& \quad = - \a \sqrt{b^2 (\la - \ul)^2 \left(P^{(n)}_\la \right)^2 + n \la \left(P^{(n)} - P^{(n-1)} \right)^2} \cr
& P^{(n)}(r, \la, T) = n.} \right. \eqno(4.1)$$

\noindent The initial value in this recursion is $P^{(0)} \equiv 0$, and the price $P$ as defined by (2.17) is $P^{(1)}$.

As in Section 2, we can multiplicatively separate the variables $r$ and $\la$ in $P^{(n)}$.  Indeed, $P^{(n)}(r, \la, t) = F(r, t) \v^{(n)}(\la, t)$, in which $F$ solves (2.5) and $\v^{(n)}$ solves the recursion

$$\left\{ \eqalign{& \v^{(n)}_t + a \v^{(n)}_\la +  {1 \over 2} b^2 (\la - \ul)^2 \v^{(n)}_{\la \la} - n \la \left(\v^{(n)} - \v^{(n-1)} \right) \cr
& \qquad = - \a \sqrt{b^2 (\la - \ul)^2 \left(\v^{(n)}_\la \right)^2 + n \la \left(\v^{(n)} - \v^{(n-1)} \right)^2}, \cr
& \v^{(n)}(\la, T) = n,} \right. \eqno(4.2)$$

\noindent with initial value $\v^{(0)} \equiv 0$.  Note that $\v$ in (2.18) equals $\v^{(1)}$.

Throughout this section, we apply Theorem 3.2 with $g = g_n$ defined by (4.3) below.  We have the following lemma whose proof we omit because it parallels that of Lemma 3.3. 

\lem{4.1} {Define $g_n,$ for $n \ge 1,$ by
$$g_n(\la, t, v, p) = a(\la, t) p - n \la \left(v - \v^{(n-1)} \right) + \a \sqrt{b^2 (\la - \ul)^2 p^2 + n \la \left(v - \v^{(n-1)} \right)^2}, \eqno(4.3)$$
in which $\v^{(n-1)}$ solves $(4.2)$ with $n$ replaced by $n-1$.  Then, $g_n$ satisfies the one-sided Lipschitz condition $(3.2)$ on $G$.  Furthermore, if $|a(\la, t)| \le K(\la - \ul) (1 + | \ln(\la-\ul)|),$ then $(3.3)$ holds.}

\medskip

\subsect{4.1. Interpreting ${1 \over n} \v^{(n)}$ as a Survival Probability}

In the first application of Theorem 3.2 and Lemma 4.1, we show that $0 \le {1\over n} \v^{(n)} \le e^{- \left(\ul - \a \sqrt{\ul} \right)(T - t)}$ for $n \ge 1$.

\th{4.2} {$0 \le \v^{(n)}(\la, t) \le n e^{- \left(\ul - \a \sqrt{\ul} \right)(T - t)}$ for $(\la, t) \in G$ and for $n \ge 0$.}

\pf  We proceed by induction to prove the upper bound.  For the ease of presentation in this proof, define $h(t) = e^{- \left(\ul - \a \sqrt{\ul} \right)(T - t)}$.  It is clear that the inequality holds for $n = 0$ because $\v^{(0)} \equiv 0$.  For $n \ge 1$, assume that $0 \le \v^{(n-1)} \le (n - 1) h(t)$, and show that $0 \le \v^{(n)} \le n h(t)$.

Define the differential operator $\cal L$ on $\cal G$ by (3.4) with $g = g_n$ from (4.3).  Because $\v^{(n)}$ solves (4.2), we have ${\cal L} \v^{(n)} = 0$.  Also,

$$\eqalign{{\cal L} n h(t) &= n(\ul - \a \sqrt{\ul}) h(t) - (n \la - \a \sqrt{n \la}) \left(n h(t) - \v^{(n-1)} \right) \cr
& \le n(\ul - \a \sqrt{\ul}) h(t) - (n \la - \a \sqrt{n \la}) (n - (n-1)) h(t) \cr
& = \left[ n(\ul - \a \sqrt{\ul}) - (n \la - \a \sqrt{n \la}) \right] h(t) \le 0.} \eqno(4.4)$$

\noindent Because ${\cal L}  n e^{- \left(\ul - \a \sqrt{\ul} \right)(T - t)} \le {\cal L} \v^{(n)}$ and $\v^{(n)}(\la, T) = n$, Theorem 3.2 and Lemma 4.1 imply that $\v^{(n)} \le n e^{- \left(\ul - \a \sqrt{\ul} \right)(T - t)}$ on $G$.

Similarly, we prove the lower bound via induction.  Suppose that $\v^{(n-1)} \ge 0$ for $n \ge 1$, and show that $\v^{(n)} \ge 0$. Let $\cal L$ be the differential operator from the first part of this proof, and denote by {\bf 0} the function that is identically 0 on $G$; then, ${\cal L} {\bf 0} = (n \la + \a \sqrt{n \la}) \v^{(n-1)} \ge 0$.  Because ${\cal L} {\bf 0} \ge {\cal L} \v^{(n)}$ and $\v^{(n)}(\la, T) = n$, Theorem 3.2 and Lemma 4.1 imply that $0 \le \v^{(n)}$ on $G$.   $\square$  

\medskip 

An immediate corollary of Theorem 4.2 is that $0 \le P^{(n)} \le n F$, as in Corollary 3.6, so, once again the price per risk ${1 \over n} P^{(n)}$ lies between 0 and $F$.  We end this subsection by extending Theorem 3.7, namely $\v_\la \le 0$, to an arbitrary number of risks.  We first present a lemma which we use in its proof.

\lem{4.3} {$\v^{(n)}(\la, t) \ge \v^{(n-1)}(\la, t)$ for $(\la, t) \in G$ and for $n \ge 1.$}

\pf  We proceed by induction.  This inequality is true for $n = 1$ because $\v^{(1)} = \v \ge 0 = \v^{(0)}$ by Theorem 3.5.  For $n \ge 2$, assume that $\v^{(n-1)} \ge \v^{(n-2)}$, and show that $\v^{(n)} \ge \v^{(n-1)}$.

Define a differential operator $\cal L$ on $\cal G$ by (3.4) with $g = g_n$ from (4.3).  Because $\v^{(n)}$ solves (4.2), we have ${\cal L} \v^{(n)} = 0$.  Also,

$$\eqalign{{\cal L} \v^{(n-1)} &= (n-1) \la \left(\v^{(n-1)} - \v^{(n-2)} \right) - n \la \left(\v^{(n-1)} - \v^{(n-1)} \right) \cr
& \quad + \a \sqrt{b^2 (\la -\ul)^2  \left(\v^{(n-1)}_\la \right)^2 + n \la \left(\v^{(n-1)} - \v^{(n-1)} \right)^2} \cr
& \quad - \a \sqrt{b^2 (\la -\ul)^2 \left(\v^{(n-1)}_\la \right)^2 + (n - 1) \la \left(\v^{(n-1)} - \v^{(n-2)} \right)^2} \cr
& \ge (n-1) \la \left(\v^{(n-1)} - \v^{(n-2)} \right) + \a b (\la - \ul) \left| \v^{(n-1)}_\la \right| \cr
& \quad - \a \left(b (\la - \ul) \left|\v^{(n-1)}_\la \right| + \sqrt{(n - 1) \la} \left(\v^{(n-1)} - \v^{(n-2)} \right) \right) \cr
& = \left( (n - 1) \la - \a \sqrt{(n - 1) \la} \right) \left(\v^{(n-1)} - \v^{(n-2)} \right) \ge 0 = {\cal L} \v^{(n)} .} \eqno(4.5)$$

\noindent Note that the first inequality follows from the subadditivity of the square root, that is, $\sqrt{A^2 + B^2} \le |A | + |B|$.  We also use the induction hypothesis that $\v^{(n-1)} \ge \v^{(n-2)}$.  In addition, $\v^{(n)}(\la, T) = n > n-1 = \v^{(n-1)}(\la, T)$.  Thus, Theorem 3.2 and Lemma 4.1 imply that $\v^{(n)} \ge \v^{(n-1)}$ on $G$.  $\square$

\medskip

Lemma 4.3 is interesting in its own right because it confirms our intuition that $P^{(n)}$ increases with the number of policyholders $n$.

\th{4.4} {$\v^{(n)}_\la(\la, t) \le 0$ for $(\la, t) \in G$ and for $n \ge 0$.}

\pf  We proceed by induction.  We know that the inequality holds when $n = 0$ and 1.  For $n \ge 2$, assume that $\v^{(n-1)}_\la \le 0$, and show that $\v^{(n)}_\la \le 0$.  As in the proof of Theorem 3.7, to prove this assertion, we apply a modified version of Theorem 3.2 to the special case of comparing $\v^{(n)}_\la$ with the zero function {\bf 0}.  First, differentiate $\v^{(n)}$'s equation with respect to $\la$ to get an equation for $f^{(n)} = \v^{(n)}_\la$.

$$\left\{ \eqalign{& f^{(n)}_t + (a_\la - n \la)f^{(n)} + n \la f^{(n-1)} + (a + b^2(\la - \ul)) f^{(n)}_\la +  {1 \over 2} b^2 (\la - \ul)^2 f^{(n)}_{\la \la} \cr
& \qquad - n \left(\v^{(n)} - \v^{(n-1)} \right) \cr
& \quad = - \a { b^2 (\la - \ul) \left(f^{(n)} \right)^2 + b^2 (\la - \ul)^2 f^{(n)} f^{(n)}_\la + {1 \over 2} n \left(\v^{(n)} - \v^{(n-1)} \right)^2  \over \sqrt{b^2 (\la - \ul)^2 \left(f^{(n)} \right)^2 + n \la \left(\v^{(n)} - \v^{(n-1)} \right)^2}} \cr
& \qquad - \a {n \la \left(\v^{(n)} - \v^{(n-1)} \right) \left(f^{(n)} - f^{(n-1)} \right) \over \sqrt{b^2 (\la - \ul)^2 \left(f^{(n)} \right)^2 + n \la \left(\v^{(n)} - \v^{(n-1)} \right)^2}}, \cr
& f^{(n)}(\la, T) = 0.} \right. \eqno(4.6)$$

Define a differential operator $\cal L$ on $\cal G$ by (3.4) with $g = g_n$ given by

$$\eqalign{&g_n(\la, t, v, p) = (a_\la - n \la)v + n \la f^{(n-1)} + (a + b^2(\la - \ul)) p - n \left(\v^{(n)} - \v^{(n-1)} \right) \cr
& + \a { b^2 (\la - \ul) v^2 + b^2 (\la - \ul)^2 v p + {n \over 2} \left(\v^{(n)} - \v^{(n-1)} \right)^2 + n \la \left(\v^{(n)} - \v^{(n-1)} \right) \left(v - f^{(n-1)} \right) \over \sqrt{b^2 (\la - \ul)^2 v^2 + n \la \left(\v^{(n)} - \v^{(n-1)} \right)^2}}.} \eqno(4.7)$$
 
\noindent From Walter (1970, Section 28, pages 213-215), we see that we only need to verify that (3.2) holds for $v > 0 = w = q$.   It is not difficult to show that Lemma 4.3 implies that $g_n$ satisfies the inequality in (3.13).  Thus, by Assumption 3.4, $g_n$ satisfies (3.2) with the corresponding $c$ and $d$ satisfying the growth conditions in (3.3).

Next, note that because $f^{(n)} = \v^{(n)}_\la$ satisfies (4.5), ${\cal L} f^{(n)} = 0$.  Also, we have ${\cal L} {\bf 0} = \left(n \la - \a \sqrt{n \la} \right) f^{(n-1)} - \left(n - \a/2 \sqrt{n/\la} \right) \left(\v^{(n)} - \v^{(n-1)} \right) \le 0$ by the induction assumption, by Lemma 4.3, and because $\la \ge \ul \ge \a^2$.  These observations, together with $f^{(n)}(\la, T) = 0$, imply that $f^{(n)} = \v^{(n)}_\la \le 0$ on $G$.  $\square$

\subsect{4.2. Comparative Statics for $P^{(n)}$}

In this section, we present properties of $P^{(n)}$ to parallel those in Section 3.2.  We show that as we vary the model parameters, the price $P^{(n)}$ responds consistently with what we expect, as did $P = P^{(1)}$ in Section 3.2.  We begin with a lemma that will help in proving Theorem 4.6 below.

\lem{4.5} {Suppose $A \ge B$ and $C$ are constants; then, $\sqrt{C^2 + A^2} \le (A - B) + \sqrt{C^2 + B^2}$.}

\pf  By squaring both sides, we see that this inequality is equivalent to $C^2 + A^2 \le (A^2 - 2AB + B^2) + (C^2 + B^2) + 2(A - B) \sqrt{C^2 + B^2}$, which simplifies to $0 \le (A - B)(\sqrt{C^2 + B^2} - B)$, which is clearly true.  $\square$

\th{4.6} {Suppose $0 \le \a_1 <  \a_2 \le \sqrt{\ul}$, and let $P^{(n), \a_i}$ be the solution to $(4.1)$ with $\a = \a_i,$ for $i = 1, 2$ and for $n \ge 0$. Then, $P^{(n), \a_1}(r, \la, t) \le P^{(n), \a_2}(r, \la, t)$ for all $(r, \la, t) \in {\bf R}^+ \times G$.}

\pf  It is enough to show that $\v^{(n), \a_1} \le \v^{(n), \a_2}$ on $G$, in which $\v^{(n), \a_i}$ has the obvious meaning.  We proceed by induction.  It is clear that the inequality holds for $n= 0$ because $\v^{0, a_i} \equiv 0$ for $i = 1, 2$.  For $n \ge 1$, assume that $\v^{(n-1), \a_1} \le \v^{(n-1), \a_2}$, and show that $\v^{(n), \a_1} \le \v^{(n), \a_2}$.

Define a differential operator $\cal L$ on $\cal G$ by (3.4) with $g = g_n$ from (4.3) with $\a = \a_1$.  Because $\v^{(n), \a_1}$ solves (4.2) with $\a = \a_1$, we have ${\cal L} \v^{(n), \a_1} = 0$.  Also,

$$\eqalign{{\cal L} \v^{(n), \a_2} &= -n \la \left(\v^{(n), \a_2} - \v^{(n-1), \a_1} \right) + n \la \left(\v^{(n), \a_2} - \v^{(n-1), \a_2} \right) \cr
& \quad + \a_1 \sqrt{b^2(\la - \ul)^2 \left(\v^{(n), \a_2}_\la \right)^2 + n \la \left(\v^{(n), \a_2} - \v^{(n-1), \a_1} \right)^2} \cr
& \quad - \a_2 \sqrt{b^2(\la - \ul)^2 \left(\v^{(n), \a_2}_\la \right)^2 + n \la \left(\v^{(n), \a_2} - \v^{(n-1), \a_2} \right)^2} \cr
& = -n \la \left(\v^{(n-1), \a_2} - \v^{(n-1), \a_1} \right) \cr
& \quad + \a_1 \left\{ \sqrt{b^2(\la - \ul)^2 \left(\v^{(n), \a_2}_\la \right)^2 + n \la \left(\v^{(n), \a_2} - \v^{(n-1), \a_1} \right)^2} \right.  \cr
& \qquad \qquad \left. - \sqrt{b^2(\la - \ul)^2 \left(\v^{(n), \a_2}_\la \right)^2 + n \la \left(\v^{(n), \a_2} - \v^{(n-1), \a_2} \right)^2} \right\} \cr
& \quad -(\a_2 - \a_1) \sqrt{b^2(\la - \ul)^2 \left(\v^{(n), \a_2}_\la \right)^2 + n \la \left(\v^{(n), \a_2} - \v^{(n-1), \a_2} \right)^2}  \cr
& \le - \left(n \la - \a_1 \sqrt{n \la} \right) \left(\v^{(n-1), \a_2} - \v^{(n-1), \a_1} \right) \le 0 = {\cal L} \v^{(n), \a_1}.} \eqno(4.8)$$

\noindent Note that the first inequality follows from Lemma 4.5 with $A = \sqrt{n \la} \left( \v^{(n), \a_2} - \v^{(n-1), \a_1} \right)$, $B = \sqrt{n \la} \left( \v^{(n), \a_2} - \v^{(n-1), \a_2} \right)$, and $C = b(\la - \ul) \v^{(n), \a_2}_\la$, from the induction hypothesis, and from $\a_2 > \a_1$.  In addition, both $\v^{(n), \a_1}$ and $\v^{(n), \a_2}$ satisfy the terminal condition $\v^{(n), \a_i}(\la, T) = 1$.  Thus, Theorem 3.2 and Lemma 4.1 imply that $\v^{(n), \a_1} \le \v^{(n), \a_2}$ on $G$.  $\square$

\medskip 

Theorem 4.6 extends Theorem 3.8 and states that as the parameter $\a$ increases, the risk-adjusted price $P^{(n), \a}$ increases.  We have the following corollary to Theorem 4.6.

\cor{4.7} {Let $P^{(n), \a0}$ be the solution to $(4.1)$ with $\a = 0;$ then, $P^{(n), \a0} \le P^{(n), \a}$ for all $0 \le \a \le \sqrt{\ul}$, and we can express the lower bound $P^{(n), \a0}$ as follows:  $P^{(n), \a0} (r, \la, t) = n F(r, t) \v^{\a0} (\la, t),$ in which $\v^{\a0}$ is given by $(3.15)$.}

\pf It is straightforward to show that $n \v^{\a0}$ solves (4.2) with $\a = 0$, and the result follows.  $\square$

\medskip

\noindent Note that $n \v^{\a0}$ is the expected number of survivors under the physical measure, so the lower bound of ${1 \over n} P^{(n)}$ (as $\a$ approaches zero) is the same as the lower bound of $P$, namely, $F \v^{\a0}$.

Next, we examine how the risk-adjusted price $P^{(n)}$ varies with the drift and volatility of the stochastic hazard rate.  We state the following two theorems without proof because their proofs extend those of Theorems 3.10 and 3.11, respectively, as the proof of Theorem 4.6 extends the one of Theorem 3.8.

\th{4.8} {Suppose $a_1(\la, t) \le a_2(\la, t)$ on $G$, and let $P^{(n), a_i}$ denote the solution to $(4.1)$ with $a = a_i,$  for $i = 1, 2$.  Then, $P^{(n), a_1}(r, \la, t) \ge P^{(n), a_2}(r, \la, t)$ for all $(r, \la, t) \in {\bf R}^+ \times G$.}

\th{4.9} {Suppose $0 \le b_1(t) \le b_2(t)$ on $[0, T],$ and let $P^{(n), b_i}$ denote the solution to $(4.1)$ with $b = b_i,$  for $i = 1, 2$.  If $P^{(n), b_i}_{\la \la} \ge 0$ for $i = 1$ or $2,$ then $P^{(n), b_1}(r, \la, t) \le P^{(n), b_2}(r, \la, t)$ for all $(r, \la, t) \in {\bf R}^+ \times G$.}

\subsect{4.3.  Subadditivity of $P^{(n)}$}

Subadditivity holds for the standard deviation premium principle, that is, $H(X) + H(Y) \ge H(X + Y)$; thus, we expect it to hold for our pricing rule.  We next show that $P^{(n)}$ is subadditive.  Specifically, we show that for $m, n$ nonnegative integers, the following inequality holds:

$$P^{(m)} + P^{(n)} \ge P^{(m+n)}.  \eqno(4.9)$$

\noindent Subadditivity is a reasonable property because if it did not hold, then buyers of insurance could  insure risks separately and thereby save money.

We begin with a lemma that we will help us prove (4.9).

\lem{4.10} {Suppose $A \ge C \ge B,$ $B_\la,$ and $C_\la$ are constants; then, for nonnegative integers $m$ and $n$,
$$\sqrt{(B_\la + C_\la)^2 + (m+n) A^2} - \sqrt{n} (A - C) \le \sqrt{B_\la^2 + m B^2} + \sqrt{C_\la^2 + n C^2} + \sqrt{m} (A - B).  \eqno{(4.10)}$$}

\pf The left-hand side of (4.10) is nonnegative because $A \ge A - C \ge 0$.  Square both sides of (4.10) to get

$$\eqalign{& (B_\la + C_\la)^2 + (m+n)A^2 + n(A - C)^2 - 2 \sqrt{n} (A - C) \sqrt{(B_\la + C_\la)^2 + (m+n) A^2} \cr
& \le B_\la^2 + m B^2 + C_\la^2 + n C^2 + m(A - B)^2 + 2 \sqrt{m} (A - B) \left\{ \sqrt{B_\la^2 + m B^2} + \sqrt{C_\la^2 + n C^2} \right\} \cr
& \quad + 2 \sqrt{B_\la^2 + m B^2} \sqrt{C_\la^2 + n C^2},}  \eqno{(4.11)}$$

\noindent which simplifies to

$$\eqalign{& B_\la C_\la + n A(A - C) + m B (A - B) \le \sqrt{n} (A - C) \sqrt{(B_\la + C_\la)^2 + (m+n) A^2} \cr
& \quad + \sqrt{m} (A - B) \left\{ \sqrt{B_\la^2 + m B^2} + \sqrt{C_\la^2 + n C^2} \right\} + \sqrt{B_\la^2 + m B^2} \sqrt{C_\la^2 + n C^2}.}  \eqno{(4.12)}$$

If the left-hand side of (4.12) is nonpositive, then we are done.  Suppose the left-hand side of (4.12) is positive, so that after squaring both sides, this inequality is equivalent to

$$\eqalign{& B_\la^2 C_\la^2 + 2 B_\la C_\la \left\{ n A(A - C) + m B(A - B) \right\} + n^2 A^2(A - C)^2 \cr
& \quad + 2 mn AB(A - B)(A - C) + m^2 B^2 (A - B)^2 \cr
& \le n (A-C)^2 \left\{ (B_\la + C_\la)^2 + (m+n) A^2 \right\} + m(A - B)^2 \left\{ B_\la^2 + C_\la^2 + m B^2 + n C^2  \right\} \cr
& \quad + (B_\la^2 + m B^2) (C_\la^2 + n C^2) + D,}  \eqno{(4.13)}$$

\noindent in which $D \ge 0$ is a sum of nonnegative square-root terms.  Inequality (4.13) simplifies to

$$\eqalign{0 &\le m \left\{ C_\la^2 (A - B)^2 + (B_\la (A - B) - C_\la B)^2 \right\} + n \left\{ B_\la^2 (A-C)^2 + (B_\la C - C_\la (A - C))^2 \right\}  \cr
& \quad + D + mn \left\{ (A(A - C) - C(A - B))^2 + B^2 C^2 + 2 A(C - B)(A - B)(A - C) \right\},}  \eqno{(4.14)}$$

\noindent which is true because $C \ge B$.  $\square$

\medskip

\th{4.11} {If $m$ and $n$ are nonnegative integers, then $P^{(m)} + P^{(n)} \ge P^{(m+n)}$.}

\pf  We prove this inequality by induction on the sum $m + n$.  We know that (4.9) holds when $m$ or $n$ equals 0 or when $m + n$ equals 0 or 1 because $P^{(0)} \equiv 0$.   For $m + n \ge 2$ with $m \ge 1$ and $n \ge 1$, assume that $P^{(k)} + P^{(\ell)} \ge P^{(k + \ell)}$ for all nonnegative integers $k$ and $\ell$ such that $k + \ell \le m + n - 1$.  We proceed to show that $P^{(m)} + P^{(n)} \ge P^{(m+n)}$.

The function $\xi = \v^{(m)} + \v^{(n)}$ solves

$$\left\{ \eqalign{& \xi_t + a \xi_\la +  {1 \over 2} b^2 (\la - \ul)^2 \xi_{\la \la} - m \la \left(\v^{(m)} - \v^{(m-1)} \right) - n \la \left(\v^{(n)} - \v^{(n-1)} \right) \cr
& \quad = - \a \sqrt{b^2 (\la - \ul)^2 \left(\v^{(m)}_\la \right)^2 + m \la \left(\v^{(m)} - \v^{(m-1)} \right)^2} \cr
& \qquad - \a \sqrt{b^2 (\la - \ul)^2 \left(\v^{(n)}_\la \right)^2 + n \la \left(\v^{(n)} - \v^{(n-1)} \right)^2}, \cr
& \xi(\la, T) = m + n,} \right. \eqno(4.15)$$

\noindent and $\phi = \v^{(m + n)}$ solves

$$\left\{ \eqalign{& \phi_t + a \phi_\la +  {1 \over 2} b^2 (\la - \ul)^2 \phi_{\la \la} - (m+n) \la \left(\phi - \v^{(m+n-1)} \right) \cr
& \quad = - \a \sqrt{b^2 (\la - \ul)^2 \phi^2_\la + (m+n) \la \left(\phi - \v^{(m+n-1)} \right)^2}, \cr
& \phi(\la, T) = m + n.} \right. \eqno(4.16)$$

Define a differential operator $\cal L$ on $\cal G$ by (3.4) with $g = g_{m+n}$ from (4.3).  From (4.16), we deduce that  ${\cal L} \phi = 0$.  Also, because $\xi$ solves (4.15), we have

$$\eqalign{{\cal L} \xi &= m \la \left(\v^{(m)} - \v^{(m-1)} \right) + n \la \left(\v^{(n)} - \v^{(n-1)} \right) - (m + n) \la \left(\xi - \v^{(m+n-1)} \right) \cr
& \quad - \a \sqrt{b^2 (\la - \ul)^2 \left(\v^{(m)}_\la \right)^2 + m \la \left(\v^{(m)} - \v^{(m-1)} \right)^2} \cr
& \quad - \a \sqrt{b^2 (\la - \ul)^2 \left(\v^{(n)}_\la \right)^2 + n \la \left(\v^{(n)} - \v^{(n-1)} \right)^2} \cr
& \quad + \a \sqrt{b^2 (\la - \ul)^2 \xi^2_\la + (m+n) \la \left(\xi - \v^{(m+n-1)} \right)^2} \cr
& \le - \left(\v^{(m)} + \v^{(n-1)} - \v^{(m+n-1)} \right) \left(n \la - \a \sqrt{n \la} \right) \cr
& \quad - \left(\v^{(m-1)} + \v^{(n)} - \v^{(m+n-1)} \right) \left(m \la - \a \sqrt{m \la} \right) \cr
& \le 0 = {\cal L} \phi.} \eqno(4.17)$$

\noindent The last inequality follows from the induction hypothesis and from $\la - \a \sqrt{\la} > 0$ for $\la > \ul \ge \a^2$.  The first inequality in (4.17) follows from Lemma 4.10 with the assignments $B_\la = b (\la - \ul) \v^{(m)}_\la$, $C_\la = b (\la - \ul) \v^{(n)}_\la$, $A = \sqrt{\la} \left(\xi - \v^{(m+n-1)} \right)$, $B = \sqrt{\la} \left(\v^{(m)} - \v^{(m-1)} \right)$, and $C = \sqrt{\la} \left(\v^{(n)} - \v^{(n-1)} \right)$.  By the induction assumption, we have $A \ge B$ and $A \ge C$; without loss of generality, $C \ge B$.  The functions $\xi$ and $\phi$ satisfy the same terminal condition when $t = T$.  Thus, Theorem 3.2 and Lemma 4.1 imply that $\xi \ge \phi$, or equivalently $P^{(m)} + P^{(n)} \ge P^{(m + n)}$.  $\square$

\medskip

Theorem 4.11 gives us another proof -- alternative to the one in Theorem 4.2 -- that $P^{(n)} \le nF$, namely, $P^{(n)} \le n P^{(1)} \le nF$, in which we use Corollary 3.6 to assert that $P^{(1)} = P \le F$.

\subsect{4.4.  Limiting Behavior of $P^{(n)}$}

We next consider the limiting behavior of $P^{(n)}$.  To motivate the results of this section, reconsider the standard deviation premium principle from (3.1) as applied to pricing $n$ risks $X_1, X_2, \dots, X_n$ that are identically distributed to a random variable $X$ and conditionally independent given the random hazard rate.  For concreteness, suppose that $X_i$ is the indicator random variable of the event that individual $i$ will be alive at time $T$.  Then, $X_S = \sum_{i=1}^n X_i$ is the total number of survivors at time $T$, so that ${\bf E} (X_S) = n {\bf E} X$, and

$$\eqalign{{\bf Var} (X_S) &= {\bf Var} [{\bf E} (X_S | {\cal S})] + {\bf E} [{\bf Var} (X_S | {\cal S})] \cr
& = n^2 {\bf Var} [{\bf E} (X | {\cal S})] + n {\bf E} [{\bf Var} (X | {\cal S})],} \eqno(4.18)$$

\noindent in which ${\cal S}$ is the $\sigma$-algebra generated by $W^\la$, the Brownian motion driving the stochastic hazard rate.  Therefore, the static standard deviation premium principle (3.1) gives us

$$H(X_S) = n {\bf E} X + \a n \sqrt{{\bf Var} [{\bf E} (X | {\cal S})] + {1 \over n} {\bf E} [{\bf Var} (X | {\cal S})]}. \eqno(4.19)$$

\noindent Note that ${1 \over n} H(X_S)$ decreases as $n$ increases.  Also, in the limit, we have

$$\lim_{n \rightarrow \infty} {1 \over n} H(X_S) = {\bf E} X + \a \sqrt{{\bf Var} [{\bf E} (X | {\cal S})]}, \eqno(4.20)$$

\noindent which is strictly greater than ${\bf E} X$ if $\a > 0$ and if ${\bf Var} [{\bf E} (X | {\cal S})] > 0$.  This last inequality will hold if $b$ is uniformly bounded below by $\kappa > 0$.  Otherwise, if $b \equiv 0$, then ${\bf Var} [{\bf E} (X | {\cal S})] = 0$, and the right-hand side of (4.20) is simply ${\bf E} X$.

In this section, we show that $P^{(n)}$ behaves much as $H(X_S)$ does in (4.19) and (4.20).  In Theorem 4.13, we show that the price per risk, ${1 \over n} P^{(n)}$, decreases as $n$ increases; that is, by increasing the number of individuals insured, we reduce the risk per individual (as measured by the price).  This result is consistent with what we expect, as inspired by (4.19).  The question answered by the results that follow Theorem 4.13 is {\it how far} does ${1 \over n} P^{(n)}$ decrease, and we obtain results in close parallel to (4.20).

We begin with a useful lemma.

\lem{4.12} {If $n \ge 2$, and if $A \ge C \ge 0$ and $B_\la$ are constants, then the following inequality holds
$$\sqrt{B_\la^2 + {1 \over n} C^2} \le \sqrt{n-2} \, (A - C) + \sqrt{B_\la^2 + {1 \over n-1} ((n-1)C - (n-2)A)^2}. \eqno(4.21)$$}

\pf By squaring both sides of (4.21), we can show that (4.21) is equivalent to

$$\eqalign{{1 \over n} C^2 &\le (n - 2)(A - C)^2 + {1 \over n - 1} ((n - 1)C - (n - 2)A)^2 \cr
& \quad + 2 \sqrt{n - 2} \, (A - C) \sqrt{B_\la^2 + {1 \over n - 1} ((n-1)C - (n-2)A)^2}.}  \eqno(4.22)$$

\noindent Because $B_\la$ is arbitrary, (4.22) is true if and only if (4.22) holds when $B_\la = 0$, which is equivalent to the following after taking the square root of the resulting right-hand side of (4.22):

$$0 \le \sqrt{n-2} \left( 1 - \sqrt{{n-2 \over n-1}} \right) (A - C) + C \left( \sqrt{n-1} - {n-2 \over \sqrt{n-1}} - {1 \over \sqrt{n}} \right). \eqno(4.23)$$

\noindent Inequality (4.23) holds if $\sqrt{n-1} - {n-2 \over \sqrt{n-1}} - {1 \over \sqrt{n}} \ge 0$, which is true for $n \ge 2$.  $\square$

\medskip

\th{4.13} {${1 \over n} P^{(n)}$ decreases with respect to $n \ge 1$.}

\pf Define $\z^{(n)} = {1 \over n} \v^{(n)}$ for $n \ge 1$.  Note that $\z^{(n)}$ solves 

$$\left\{ \eqalign{& \z^{(n)}_t + a \z^{(n)}_\la +  {1 \over 2} b^2 (\la - \ul)^2 \z^{(n)}_{\la \la} - \la \left(n \z^{(n)} - (n - 1) \z^{(n-1)} \right) \cr
& \qquad = - \a \sqrt{b^2 (\la - \ul)^2  \left(\z^{(n)}_\la \right)^2 + {1 \over n} \la \left(n \z^{(n)} - (n-1) \z^{(n-1)} \right)^2}, \cr
& \z^{(n)}(\la, T) = 1,} \right. \eqno(4.24)$$

\noindent with $\z^{(1)} = \v^{(1)} = \v$.

We proceed by induction and show that $\z^{(n)} \le \z^{(n-1)}$ for $n \ge 2$.  We first show that $\z^{(2)} \le \z^{(1)} = \v$.  Define a differential operator ${\cal L}$ on $\cal G$ by (3.4) with $g = g_2$ given by

$$g_2(\la, t, v, p) = a(\la, t) p - \la (2v - \v) + \a \sqrt{b^2 (\la - \ul)^2 p^2 + {1 \over 2} \la (2v - \v)^2}.  \eqno(4.25)$$

\noindent Clearly, $g_2$ satisfies (3.2) and (3.3), so we can apply Theorem 3.2. Note that because $\z^{(2)}$ solves (4.24) with $n = 2$, ${\cal L} \z^{(2)} = 0$.  Also,

$${\cal L} \v = \a \left\{ \sqrt{b^2 (\la - \ul)^2 \v_\la^2 + {1 \over 2} \la \v^2} - \sqrt{b^2 (\la - \ul)^2 \v_\la^2 + \la \v^2} \right\} \le 0 = {\cal L} \z^{(2)}.  \eqno(4.26)$$

\noindent Additionally, $\z^{(2)}(\la, T) = 1 = \v(\la, T)$; thus, $\z^{(2)} \le \z^{(1)} = \v$.

Next, assume that for $n \ge 3$, $\z^{(n-1)} \le \z^{(n-2)}$, and show that $\z^{(n)} \le \z^{(n-1)}$.  Define a differential operator $\cal D$ on $\cal G$ by (3.4) with $g = g_n$ given by

$$g_n(\la, t, v, p) = a p - \la \left(n v - (n - 1) \z^{(n-1)} \right) + \a \sqrt{b^2 (\la - \ul)^2 p^2 + {1 \over n} \la \left(n v - (n-1) \z^{(n-1)} \right)^2}.  \eqno(4.27)$$

\noindent Clearly, $g_n$ satisfies (3.2) and (3.3), so we can apply Theorem 3.2. Note that because $\z^{(n)}$ solves (4.24), ${\cal D} \z^{(n)} = 0$.  Also,

$$\eqalign{{\cal D} \z^{(n-1)} &= (n-2) \la \left(\z^{(n-1)} - \z^{(n-2)} \right) \cr
& \quad - \a \left\{ \sqrt{b^2 (\la - \ul)^2 \left(\z^{(n-1)}_\la \right)^2 + {1 \over n-1} \la \left((n-1) \z^{(n-1)} - (n-2) \z^{(n-2)} \right)^2} \right. \cr
& \qquad \qquad \left. - \sqrt{b^2 (\la - \ul)^2 \left(\z^{(n-1)}_\la \right)^2 + {1 \over n} \la \left(\z^{(n-1)} \right)^2} \right\} \cr
& \le \left((n-2) \la - \a \sqrt{(n-2) \la} \right) \left(\z^{(n-1)} - \z^{(n-2)} \right) \le 0 = {\cal D} \z^{(n)}.}  \eqno(4.28)$$

\noindent The first inequality in (4.28) follows from Lemma 4.12 under the assignments $A = \break \sqrt{\la} \z^{(n-2)}$, $C = \sqrt{\la} \z^{(n-1)}$, and $B_\la = b(\la - \ul) \z^{(n-1)}_\la$.  We also use the induction assumption.  Additionally, $\z^{(n)}(\la, T) = 1 = \z^{(n-1)}(\la, T)$; thus, $\z^{(n)} \le \z^{(n-1)}$ on $G$.  $\square$

\medskip

In the next two theorems, we answer the question motivated by Theorem 4.13, that is, we determine the limiting value of the decreasing sequence ${1 \over n} P^{(n)}$.  First, we show that ${1 \over n} P^{(n)}$ is bounded below by  $F \b$, in which $\b$ solves

$$\left\{ \eqalign{& \b_t + (a - \a b (\la - \ul)) \b_\la +  {1 \over 2} b^2 (\la - \ul)^2 \b_{\la \la} - \la \b = 0, \cr
& \b(\la, T) = 1.} \right. \eqno(4.29)$$

\noindent  Intuitively, the function $\beta$ is less than $\v$ because we replaced the square root in $\v$'s pde with the square root of the first term.  Later, we show that ${1 \over n} P^{(n)}$ equals $F \b$ in the limit.  In other words, $\lim_{n \rightarrow \infty} {1 \over n} \v^{(n)} = \b$.

In order to prove that ${1 \over n} P^{(n)} \ge F \b$, we require the following lemma.

\lem{4.14} {The function $\b$ given by $(4.29)$ is nonincreasing with respect to $\la$.}

\pf Differentiate $\b$'s equation with respect to $\la$ to get an equation for $f = \b_\la$.  The function $f$ solves

$$\left\{ \eqalign{& f_t + (a_\la - \a b - \la) f + (a - \a b (\la - \ul) + b^2 (\la - \ul)) f_\la +  {1 \over 2} b^2 (\la - \ul)^2 f_{\la \la} - \b = 0, \cr
& f(\la, T) = 0.} \right. \eqno(4.30)$$

\noindent Thus, $f$'s equation is of the form in (3.4) with $g$ given by

$$g(\la, t, v, p) = (a_\la - \a b - \la) v + (a - \a b (\la - \ul) + b^2 (\la - \ul)) p - \b.  \eqno(4.31)$$

Because of the growth condition assumed for $a_\la$ in Assumption 3.4, it is straightforward to show that $g$ satisfies the one-sided Lipschitz condition in (3.2) with growth conditions in (3.3).  Define a differential operator $\cal L$ on $\cal G$ by (3.4) with $g$ given by (4.31).  Because $f$ solves (4.30), we have ${\cal L} f = 0$.  Denote by {\bf 0} the function that is identically 0 on $G$; then, ${\cal L} {\bf 0} = - \b \le 0 = {\cal L} f$.  In addition, $f(\la, T) = 0$; thus, Theorem 3.2 implies that $f = \b_\la \le 0$.  $\square$   

\medskip

\th{4.15} {$$\lim_{n \rightarrow \infty}{1 \over n} P^{(n)}(r, \la, t) \ge F(r, t) \b(\la, t). \eqno(4.32)$$}

\pf  It is enough to show that the solution $\b$ of $(4.29)$ is a lower bound of ${1 \over n} \v^{(n)}$.  We proceed by induction.  First, show that $\b \le \v = \v^{(1)}$.  Define a differential operator $\cal L$ on $\cal G$ by (3.4) with $g$ given by (3.8).  Because $\v$ solves (2.18), ${\cal L} \v = 0$.  Also,

$${\cal L} \b = \a \left\{ \sqrt{b^2 (\la - \ul)^2 \b^2_\la + \la \b^2} - b (\la - \ul) \big| \b_\la \big| \right\} \ge 0 = {\cal L} \v.  \eqno(4.33)$$

\noindent In addition, $\b(\la, T) = \v(\la, T) = 1$; thus, Theorem 3.2 and Lemma 3.3 imply that $\b \le \v$.

Next, assume that $\b \le {1 \over n - 1} \v^{(n-1)}$, and show that $\b \le {1 \over n} \v^{(n)}$.  Recall that the function $\z^{(n)} = {1 \over n} \v^{(n)}$ solves (4.24).  Define a differential operator $\cal D$ on $\cal G$ by (3.4) with $g = g_n$ given in (4.27).  Recall that $(n - 1) \z^{(n-1)} = \v^{(n-1)}$ in (4.27).  Then, ${\cal D} \z^{(n)} = 0$, and

$$\eqalign{{\cal D} \b &= \b_t + a \b_\la +  {1 \over 2} b^2 (\la - \ul)^2 \b_{\la \la} - \la \left(n \b - \v^{(n-1)} \right) \cr
& \quad + \a \sqrt{b^2 (\la - \ul)^2 \b_\la^2 + {1 \over n} \la \left(n \b - \v^{(n-1)} \right)^2} \cr
& = \la \left(\v^{(n-1)} - (n - 1) \b \right) + \a \left\{ \sqrt{b^2 (\la - \ul)^2 \b_\la^2 + {1 \over n} \la \left(n \b - \v^{(n-1)} \right)^2} - b (\la - \ul) \big| \b_\la \big| \right\} \cr
& \ge 0 = {\cal D} \z^{(n)}.} \eqno(4.34)$$

\noindent Also, $\z^{(n)}(\la, T) = \b(\la, T) = 1$; thus, Theorem 3.2 and Lemma 4.1 imply that $\b \le \z^{(n)} = {1 \over n} \v^{(n)}$.  $\square$

\medskip

The next theorem tightens the result of Theorem 4.15 and shows that we have equality in (4.32).  Consider the solution to the following pde:

$$\left\{ \eqalign{& \g^{(n)}_t + (a - \a b (\la - \ul)) \g^{(n)}_\la +  {1 \over 2} b^2 (\la - \ul)^2 \g^{(n)}_{\la \la} - \left(n \la - \a \sqrt{n \la} \right) \left(\g^{(n)} - \g^{(n-1)} \right) = 0, \cr
& \g^{(n)}(\la, T) = n,} \right. \eqno(4.35)$$

\noindent in which $\g^{(0)} \equiv 0$.  We proceed by showing in a series of lemmas that $\g^{(n)} \ge \v^{(n)}$ for $n \ge 0$.  Intuitively, $\g^{(n)}$ is greater than $\v^{(n)}$ because we replaced the square root in $\v^{(n)}$'s pde with the sum of the square roots of the two terms.  Finally, we show that ${1 \over n} \g^{(n)} - \b$ goes to zero as $n$ goes to infinity.

\lem{4.16} {The function $\g^{(n)}$ given by $(4.35)$ is nonincreasing with respect to $\la$ and $\g^{(n+1)} \ge \g^{(n)}$ for $n \ge 0$.}

\pf  The proof of that $\g^{(n)}_\la \le 0$ is similar to the proof that $\v^{(n)}_\la \le 0$ in Theorems 3.7 and 4.4, so we omit the details.  Similarly, $\g^{(n+1)} \ge \g^{(n)}$ follows as in the proof of Lemma 4.3.  $\square$

\medskip




\lem{4.17} {$\g^{(n)} \ge \v^{(n)}$ for $n \ge 0$.}

\pf The result is true for $n = 0$ because $\g^{(0)} = \v^{(0)} = 0$. Suppose for $n \ge 1$, we have $\g^{(n-1)} \ge \v^{(n-1)}$, and show that $\g^{(n)} \ge \v^{(n)}$.  Define a differential operator ${\cal L}$ on $\cal G$ by (3.4) with $g = g_n$ given in (4.3).  Then, ${\cal L} \v^{(n)} = 0$, and

$$\eqalign{{\cal L} \g^{(n)} &= \a b (\la - \ul) \g^{(n)}_\la - n \la \left(\g^{(n)} - \v^{(n-1)} \right) + \left(n \la - \a \sqrt{n \la} \right) \left(\g^{(n)} - \g^{(n-1)} \right) \cr
& \quad + \a \sqrt{b^2 (\la - \ul)^2 \left(\g^{(n)}_\la \right)^2 + n \la \left(\g^{(n)} - \v^{(n-1)} \right)^2} \cr
& \le \left(n \la - \a \sqrt{n \la} \right) \left(\v^{(n-1)} - \g^{(n-1)} \right) \le 0 = {\cal L} \v^{(n)}.}  \eqno(4.36)$$

\noindent In the first inequality in (4.36), we use the facts that $\g^{(n)}_\la \le 0$ and $\g^{(n)} \ge \g^{(n-1)}$, and we use the subadditivity of the square root function.  In the second inequality, we use the induction assumption.  We also have the terminal conditions $\g^{(n)} (\la, T) = n = \v^{(n)}(\la, T)$; thus, Theorem 3.2 and Lemma 4.1 imply that $\g^{(n)} \ge \v^{(n)}$.  $\square$

\medskip

Now, we are ready to prove the main theorem of this section.

\th{4.18} {$$\lim_{n \rightarrow \infty}{1 \over n} P^{(n)}(r, \la, t) = F(r, t) \b(\la, t). \eqno(4.37)$$}

\pf  By Theorem 4.15 and Lemma 4.17, the theorem is proved if we show that ${1 \over n} \g^{(n)} - \b$ goes to zero as $n$ goes to infinity because ${1 \over n} \g^{(n)} - \b \ge {1 \over n} \v^{(n)} - \b \ge 0$.

Define $\P^{(n)}$ on $G$ by $\P^{(n)} = {1 \over n} \g^{(n)} - \b$, so the theorem is proved if we show that $\lim_{n \rightarrow \infty} \P^{(n)}(\la, t) = 0$.  The function $\P^{(n)}$ solves the recursion

$$\left\{ \eqalign{& \P^{(n)}_t + (a - \a b(\la - \ul)) \P^{(n)}_\la + {1 \over 2} b^2 (\la - \ul)^2 \P^{(n)}_{\la \la} - \left(n \la - \a \sqrt{n \la} \right) \P^{(n)} \cr
& \quad = - \a \sqrt{{\la \over n}} \b - (n-1) \left(\la - \a \sqrt{{\la \over n}} \right) \P^{(n-1)}, \cr
&\P^{(n)}(\la, T) = 0,} \right. \eqno(4.38)$$

\noindent with $0 \le \P^{(1)} = \g^{(1)} - \b \le 1$.  From (4.38) and the Feynman-Kac Theorem, we deduce the following expression for $\Phi^{(n)}$ in terms of $\Phi^{(n-1)}$:

$$\eqalign{\P^{(n)}(\la, t) &=  \a {\bf \tilde E} \left[ \int_t^T \sqrt{{ \la_s \over n}} \, \b( \la_s, s) e^{-\int_t^s (n  \la_u - \a \sqrt{n  \la_u}) du} ds \Bigg|  \la_t = \la  \right] \cr
& + (n - 1) {\bf \tilde E} \left[ \int_t^T \left(  \la_s - \a \sqrt{{ \la_s \over n}} \right) \P^{(n-1)}( \la_s, s)  e^{-\int_t^s (n  \la_u - \a \sqrt{n  \la_u}) du} ds  \Bigg|  \la_t = \la \right],} \eqno(4.39)$$

\noindent in which $\la$ follows the process $d \la_s = (a - \a b(\la_s - \ul)) ds + b (\la_s - \ul) d \tilde W^\la_s$, with $\tilde W^\la_s = W^\la_s + \a s$.  The process $\tilde W^\la$ is a standard Brownian motion with respect to the probability space $(\Omega, {\cal F}, {\bf \tilde P})$, in which ${d{\bf \tilde P} \over d{\bf P}} = e^{-\a W_T - {1 \over 2} \a^2 T}$.  $\bf \tilde E$ denotes expectation with respect to $\bf \tilde P$.  Note that $\a$ is analogous to the bond market's price of risk $q$ in (2.4).

Suppose $\P^{(n-1)}(\la, t) \le K_{n - 1}$ on $G$ for some $n \ge 2$.  Then, we can bound  $\P^{(n)}$ as follows:

$$\eqalign{\P^{(n)}(\la, t) &\le  \a {\bf \tilde E} \left[ \int_t^T \sqrt{{ \la_s \over n}} e^{-\int_t^s (n  \la_u - \a \sqrt{n  \la_u}) du} ds \Bigg|  \la_t = \la  \right] \cr
& \quad+ (n - 1) K_{n-1} {\bf \tilde E} \left[ \int_t^T \left(  \la_s - \a \sqrt{{ \la_s \over n}} \right)  e^{-\int_t^s (n  \la_u - \a \sqrt{n  \la_u}) du} ds  \Bigg|  \la_t = \la \right],} \eqno(4.40)$$

\noindent in which we use the fact that $0 \le \b \le 1$.  Define $f^{(n)}$ and $h^{(n)}$ for $n \ge 2$ by

$$f^{(n)}(\la, t) = \a {\bf \tilde E} \left[ \int_t^T n \sqrt{ \la_s} e^{-\int_t^s (n  \la_u - \a \sqrt{n  \la_u}) du} ds \Bigg|  \la_t = \la  \right], \eqno(4.41)$$

\noindent and

$$h^{(n)}(\la, t) = {\bf \tilde E} \left[ \int_t^T \left(n  \la_s - \a \sqrt{n  \la_s} \right)  e^{-\int_t^s (n  \la_u - \a \sqrt{n  \la_u}) du} ds  \Bigg|  \la_t = \la \right]. \eqno(4.42)$$

\noindent Thus, inequality (4.40) is equivalent to

$$\P^{(n)}(\la, t) \le {1 \over n^{3/2}} f^{(n)}(\la, t) + {n - 1 \over n} K_{n-1} h^{(n)}(\la, t).  \eqno(4.43)$$

After proving the following two lemmas that give us bounds on $f^{(n)}$ and $h^{(n)}$ in (4.41) and (4.42), respectively, we finish the proof of Theorem 4.18.

\lem{4.19} {The function $f^{(n)}$ defined by $(4.41)$ is bounded above by $J = {\a \sqrt{2} \over \sqrt{2 \ul} - \a}$  for all $n \ge 2$.}

\pf  By the Feynman-Kac Theorem, $f^{(n)}$ solves the linear pde

$$\left\{ \eqalign{& f^{(n)}_t + (a - \a b(\la - \ul)) f^{(n)}_\la + {1 \over 2} b^2 (\la - \ul)^2 f^{(n)}_{\la \la} - (n \la - \a \sqrt{n \la}) f^{(n)} = - \a n \sqrt{\la}, \cr
&f^{(n)}(\la, T) = 0.} \right. \eqno(4.44)$$

\noindent Define a differential operator $\cal L$ on $\cal G$ via $f^{(n)}$'s equation.  Let $\bf J$ denote the function that is identically equal to $J$.  Thus, ${\cal L}f^{(n)} = 0$, and ${\cal L}{\bf J} = - \left(n \la - \a \sqrt{n \la} \right) J + \a n \sqrt{\la} \le 0 = {\cal L}f^{(n)}$ for $n \ge 2$.  We also have $f^{(n)}(\la, T) = 0 \le J$; thus, from Theorem 3.2, we conclude that $f^{(n)} \le J$ on $G$ for $n \ge 2$.  $\square$

\medskip

\lem{4.20} {The function $h^{(n)}$ defined by $(4.42)$ is bounded above by 1 for all $n \ge 2$.}

\pf By the Feynman-Kac Theorem, $h^{(n)}$ solves the linear pde

$$\left\{ \eqalign{& h^{(n)}_t + (a - \a b(\la - \ul)) h^{(n)}_\la + {1 \over 2} b^2 (\la - \ul)^2 h^{(n)}_{\la \la} - (n \la - \a \sqrt{n \la} ) h^{(n)} = - (n \la - \a \sqrt{n \la} ), \cr
&h^{(n)}(\la, T) = 0.} \right. \eqno(4.45)$$

\noindent Define a differential operator $\cal L$ on $\cal G$ via $h^{(n)}$'s equation; thus, ${\cal L}h^{(n)} = 0 = {\cal L}{\bf 1}$.  We also have $h^{(n)}(\la, T) = 0 \le 1$; thus, from Theorem 3.2, we conclude that $h^{(n)} \le 1$ on $G$ for $n \ge 2$.  $\square$

\medskip

\noindent {\it End of Proof of Theorem 4.18.} We have shown that if $\P^{(n-1)} \le K_{n-1}$, then $\P^{(n)} \le K_n$, in which for $n \ge 2$,

$$K_n = {J \over n^{3/2}} + {n - 1 \over n} K_{n - 1}, \eqno(4.46)$$

\noindent with $K_1 = 1$.  In hindsight, we could have used induction to show that if $\P^{(n-1)} \le K_{n-1}$, then $\P^{(n)} \le K_n$ via a comparison argument with a differential operator based on the pde in (4.38).  However, we have chosen to leave the proof as is because of its constructive nature in demonstrating the origin of (4.46).

Define $L_n = n K_n$; thus,

$$L_n = L_{n - 1} + {J \over \sqrt{n}}, \quad n \ge 2, \eqno(4.47)$$

\noindent from which it follows that

$$L_n = 1 + \sum_{i = 2}^n {J \over \sqrt{i}} \le 1 + J \int_1^n {dx \over \sqrt{x}} \le 1 + 2J \sqrt{n}, \quad n \ge 2. \eqno(4.48)$$

\noindent Finally, we have

$$\P^{(n)}(\la, t) \le K_n \le {1 \over n} + {2 J \over \sqrt{n}}, \quad n \ge 1, \eqno(4.49)$$

\noindent The right-hand side of inequality (4.49) goes to zero as $n$ goes to $\infty$; thus, $\P^{(n)}(\la, t)$ goes to zero as $n$ goes to $\infty$.  In other words, $\lim_{n \rightarrow \infty} {1 \over n} P^{(n)}(\la, r, t) = F(r, t) \b(\la, t)$, as we wished to show.  $\square$

\medskip

We have the following corollaries of Theorem 4.18.  The first parallels the observation that the right-hand side in (4.20) is simply ${\bf E} X$ if $b \equiv 0$. 

\cor{4.21} {If $b \equiv 0,$ then $\lim_{n \rightarrow \infty} {1 \over n} P^{(n)}(\la, r, t) = F(r, t) \, \v^{\a0}$ in which $\v^{\a0}$ is the physical probability of survival given in $(3.15)$.}

\pf From Theorem 4.18, we know that ${1 \over n} \v^{(n)}$ goes to $\b$ as $n$ goes to infinity.  Therefore, the corollary follows because $\b = \v^{\a0}$ when $b \equiv 0$, which is clear from (4.29) by setting $b$ equal to 0.  $\square$

\medskip

\cor{4.22} {If $b$ is uniformly bounded below by $\kappa > 0$, then $\lim_{n \rightarrow \infty} {1 \over n} P^{(n)}(\la, r, t) \ge F(r, t) \, \v^{\a0}$, with equality only when $t = T$.}

\pf This result follows from the fact that $\b \ge \v^{\a0},$ with equality only when $t = T$.  Indeed, define a differential operator $\cal L$ on $\cal G$ by (3.4) and (3.8) with $\a = 0$.  Thus, ${\cal L} \v^{\a0} = 0$, and

$${\cal L} \b = \a b(\la - \ul) \b_\la \le 0 = {\cal L} \v^{\a0}.  \eqno(4.50)$$

\noindent Also, $\v^{\a0}(\la, T) = \b(\la, T) = 1$; thus, Theorem 3.2 and Lemma 3.3 imply that $\v^{\a0} \le \b$.

To show that $\v^{\a0} < \b$ for $t \in [0, T)$, first consider the pde of $\v^{\a0}_\la$.  The function $f = \v^{\a0}_\la$ solves

$$\left\{ \eqalign{& f_t + (a_\la - \la) f + (a + b^2 (\la - \ul)) f_\la +  {1 \over 2} b^2 (\la - \ul)^2 f_{\la \la} - \v^{\a0} = 0, \cr
& f(\la, T) = 0.} \right. \eqno(4.51)$$

\noindent From the linear pde in (4.51) and the Feynman-Kac Theorem, we deduce that

$$\v^{\a0}_\la(\la, t) = - {\bf \hat E} \left[ \int_t^T \v^{\a0}( \la_s, s) e^{- \int_t^s ( \la_u - a_\la( \la_u, u)) du} ds \, \Bigg| \,  \la_s = \la \right], \eqno(4.52)$$

\noindent in which $ \la$ follows the diffusion $d \la_s = (a + b^2(\la_s - \ul)) ds + b(\la_s - \ul) d \hat W^\la_s$, with $\hat W^\la_s = W^\la_s - \int_0^s b(u) du$.  The process $\hat W^\la$ is a standard Brownian motion with respect to a suitably-defined probability space, and $\bf \hat E$ denotes expectation on that space.  Note that $\v^{\a0}_\la(\la, t) < 0$ for $\la > \ul$ and for $t \in [0, T)$ because $\v^{\a0}(\la, t) > 0$ on that domain by the representation in (3.15).

Next, consider the pde of $B = \b - \v^{\a0}$.

$$\left\{ \eqalign{& B_t + (a - \a b (\la - \ul)) B_\la +  {1 \over 2} b^2 (\la - \ul)^2 B_{\la \la} - \la B = \a b(\la - \ul) \v^{\a0}_\la, \cr
& B(\la, T) = 0.} \right. \eqno(4.53)$$

\noindent From the linear pde in (4.53) and the Feynman-Kac Theorem, we deduce that

$$B(\la, t) = - \a {\bf \tilde E} \left[ \int_t^T b(s) ( \la_s - \ul) \v^{\a0}_\la( \la_s, s) e^{- \int_t^s  \la_u du} ds \, \Bigg| \,  \la_s = \la \right], \eqno(4.54)$$

\noindent in which $\la$ follows the diffusion from the paragraph following (4.39) and $\bf \tilde E$ is as in (4.39).  Note that $B(\la, t) > 0$ for $\la > \ul$ and for $t \in [0, T)$ because $\v^{\a0}_\la(\la, t) < 0$ on that domain by the representation in (4.52).  $\square$

\medskip

After all this work, we can finally decompose the risk charge $P - P^{\a0}$ -- first mentioned in the Introduction, then briefly in Section 2.2, and again following Corollary 3.9 -- into its component risk charges:  one for a finite portfolio and another for stochastic mortality.  More generally, we decompose the per-risk risk charge ${1 \over n} P^{(n)} - P^{\a0}$ when the insurer holds a portfolio of $n$ risks.  Recall from Theorem 4.18 that $\lim_{n \rightarrow \infty} {1 \over n} P^{(n)} = F \b$.  Therefore, define ${1 \over n} P^{(n)} - F \b$ as the risk charge (per risk) for holding a finite portfolio, and  define $F \b - P^{\a0}$ as the risk charge for stochastic mortality even after selling to an arbitrarily large group.  Thus, we have

$${1 \over n} P^{(n)} - P^{\a0} = \left({1 \over n} P^{(n)} - F \b \right) + \left(F \b - P^{\a0} \right) = F \left({1 \over n} \v^{(n)} - \b \right) + F  \left(\b - \v^{\a0} \right), \eqno(4.55)$$

\noindent in which the risk charge for stochastic mortality, namely $F(\b - \v^{\a0})$, is zero if $b \equiv 0$ by Corollary 4.21 and is positive (for $t < T$) if $b \ge \kappa > 0$ by Corollary 4.22.

\sect{5. Summary and Conclusions}

We developed a theoretical foundation for valuing mortality risk by assuming that the risk is ``priced'' via the instantaneous Sharpe ratio.  Because the market for pure endowments is incomplete, one cannot assert that there is a unique price.  However, we believe that the price that our method produces is a valid one because of the many desirable properties that it satisfies.  In particular, we studied properties of the price for $n$ conditionally independent and identically distributed pure endowment risks.  In Theorem 4.11, we showed that the price is subadditive with respect to $n$, and in Theorem 4.13, we showed that the risk charge per person decreases as $n$ increases.  We also proved that if the hazard rate is deterministic, then the risk charge per person goes to zero as $n$ goes to infinity (Theorem 4.20 and Corollary 4.21).  Moreover, we proved that if the hazard rate is stochastic, then the risk charge person is positive as $n$ goes to infinity, which reflects the fact that the mortality risk is not diversifiable in this case (Theorem 4.20 and Corollary 4.22).  Additionally, in equation (4.55), we decomposed the per-risk risk charge into the finite portfolio and stochastic mortality risk charges.  Because of these properties, we anticipate that our pricing methodology will prove useful in pricing risks in other incomplete markets.

In addition to addressing the problem of the breakdown in the law of large numbers, the study of such a mortality risk premium might breathe new life into the analysis of exotic options that are embedded within insurance and pension contracts, which have traditionally been viewed as being out-of-the-money and, hence, valueless.  Indeed, our theoretical framework -- which incorporates a risk measure for mortality risk -- could be used to evaluate the risks inherent in a number of recent industry trends, namely (i) the emergence of longevity-linked bonds, (ii) capacity constraints in the immediate annuity market, and (iii) the decline of defined benefit pension plans.  

\sect{Acknowledgements}

We thank Erhan Bayraktar, Kristen S. Moore, Jeffrey Rauch, Keith Promislow, and Zhengfang Zhou for their valuable help.

\sect{References}

\noindent \hangindent 20 pt Biffis, E. (2005), Affine processes for dynamic mortality and actuarial valuation, {\it Insurance: Mathematics and Economics}, to appear.

\smallskip \noindent \hangindent 20 pt Bj\"ork, T. (1998), {\it Arbitrage Theory in Continuous Time}, Oxford University Press, Oxford.

\smallskip \noindent \hangindent 20 pt Blanchet-Scalliet, C., N. El Karoui, and L. Martellini (2005), Dynamic asset pricing theory with uncertain time-horizon, {\it Journal of Economic Dynamics and Control}, 29: 1737-1764.

\smallskip \noindent \hangindent 20 pt Boyle, P. P. and M. Hardy (2003), Guaranteed annuity options, {\it ASTIN Bulletin}, 33: 125-152.

\smallskip \noindent \hangindent 20 pt Brennan, M. J. and E. Schwartz (1976), The pricing of equity-linked life
insurance policies with an asset value guarantee, {\it Journal of Financial Economics}, 3 (1): 195-213.


\smallskip \noindent \hangindent 20 pt Cairns, A. J. G., D. Blake and K. Dowd (2004), Pricing framework for
securitization of mortality risk, working paper, Heriot-Watt University.

\smallskip \noindent \hangindent 20 pt Cox, S. H. and Y. Lin (2004), Natural hedging of life and annuity mortality risks, {\it Journal of Risk and Insurance}, to appear.

\smallskip \noindent \hangindent 20 pt Dahl, M. (2004), Stochastic mortality in life insurance: Market reserves and
mortality-linked insurance contracts, {\it Insurance: Mathematics and Economics}, 35: 113-136.

\smallskip \noindent \hangindent 20 pt DiLorenzo, E. and M. Sibillo (2003), Longevity risk: Measurement and
application perspectives, working paper, Universita degli Studi di Napoli.

\smallskip \noindent \hangindent 20 pt Gerber, H. U. (1979), {\it Introduction to Mathematical Risk Theory}, Huebner Foundation Monograph 8, Wharton School of the University of Pennsylvania, Richard D. Irwin, Homewood, IL.

\smallskip \noindent \hangindent 20 pt Gerber, H. U. and E. S. W. Shiu (1994), Option pricing by Esscher transforms (with discussions), {\it Transactions of the Society of Actuaries}, 46: 99-191.

\smallskip \noindent \hangindent 20 pt Karatzas, I. and S. E. Shreve (1991), {\it Brownian Motion and Stochastic Calculus}, second edition, Springer-Verlag, New York.

\smallskip \noindent \hangindent 20 pt Lamberton, D. and B. Lapeyre (1996), {\it Introduction to Stochastic Calculus Applied to Finance}, Chapman \& Hall/CRC, Boca Raton, Florida.

\smallskip \noindent \hangindent 20 pt Lee, R. D. and L. R. Carter (1992), Modeling and forecasting U.S. mortality, {\it Journal of the American Statistical Association}, 87 (419): 659-671.

\smallskip \noindent \hangindent 20 pt Milevsky, M. A. and S. D. Promislow (2001), Mortality derivatives and the option to annuitize, {\it Insurance: Mathematics and Economics}, 29: 299-318.

\smallskip \noindent \hangindent 20 pt Norberg, R. (2004), Vasicek beyond the normal, {\it Mathematical Finance}, 14 (4): 585-604.

\smallskip \noindent \hangindent 20 pt Olivieri, A. (2001), Uncertainty in mortality projections: An actuarial
perspective, {\it Insurance: Mathematics and Economics}, 29: 231-245.

\smallskip \noindent \hangindent 20 pt Protter, P. (1995), {\it Stochastic Integration and Differential Equations}, Applications in Mathematics, 21, Springer-Verlag, Berlin.

\smallskip \noindent \hangindent 20 pt Royden, H. L. (1968), {\it Real Analysis}, second edition, Macmillan, New York.

\smallskip \noindent \hangindent 20 pt Schrager, D.F. (2005), Affine stochastic mortality, {\it Insurance:
Mathematics and Economics}, to appear.

\smallskip \noindent \hangindent 20 pt Smith, A., I. Moran, and D. Walczak (2003), Why can financial firms charge for diversifiable risk?, working paper, Deloitte Touche Tohmatsu.

\smallskip \noindent \hangindent 20 pt Soininen, P. (1995), Stochastic variation of interest and mortality, {\it Proceedings of the 5th AFIR International Colloquium}, 871-904.

\smallskip \noindent \hangindent 20 pt Walter, W. (1970), {\it Differential and Integral Inequalities}, Springer-Verlag, New York.

\smallskip \noindent \hangindent 20 pt Windcliff, H., J. Wang, P. A. Forsyth, and K. R. Vetzal (2005), Hedging with a correlated asset: Solution of a nonlinear pricing pde, working paper, University of Waterloo.

\smallskip \noindent \hangindent 20 pt Zariphopoulou, T., (2001), Stochastic control methods in asset pricing, {\it Handbook of \break Stochastic Analysis and Applications}, D. Kannan and V. Lakshmikantham (editors), Marcel Dekker, New York.

\bye